\documentclass[12pt]{article}

\usepackage{amsmath}
\usepackage{amssymb}
\allowdisplaybreaks
\usepackage{epsf}

\begin{document}
\baselineskip=14.5pt plus 0.2pt minus 0.1pt
\renewcommand{\theequation}{\thesection.\arabic{equation}}
\renewcommand{\thefootnote}{\fnsymbol{footnote}}
\makeatletter
\@addtoreset{equation}{section}
\makeatother

\newcommand{\A}{W}
\newcommand{\wt}{\widetilde}
\newcommand{\wh}{\widehat}
\newcommand{\ol}{\overline}
\newcommand{\nn}{\nonumber}
\newcommand{\half}{\frac{1}{2}}
\newcommand{\tr}{\mathop{\rm tr}\nolimits}
\newcommand{\Tr}{\mathop{\rm Tr}\nolimits}
\newcommand{\cO}{{\mathcal O}}
\newcommand{\cA}{{\mathcal A}}
\newcommand{\cF}{{\mathcal F}}
\newcommand{\bR}{\mathbb{R}}
\newcommand{\bZ}{\mathbb{Z}}
\newcommand{\cl}{{\rm cl}}
\newcommand{\Ah}{\widehat{A}}
\newcommand{\Fh}{\widehat{F}}
\newcommand{\ds}{\displaystyle}
\newcommand{\bm}[1]{\boldsymbol{#1}}


\newcommand{\SCS}{S_{\rm CS}}
\newcommand{\newSCS}{S^{\rm new}_{\rm CS}}
\newcommand{\SYM}{S_{\rm YM}}
\newcommand{\p}{\partial}
\newcommand{\Pmatrix}[1]{\begin{pmatrix} #1 \end{pmatrix}}
\newcommand{\sPmatrix}[1]{
            \left(\begin{smallmatrix} #1 \end{smallmatrix}\right)}

\newcommand{\cP}{{\mathcal P}}
\newcommand{\sd}{\wt{d}}
\newcommand{\Drv}[2]{\frac{\p #1}{\p #2}}
\newcommand{\drv}[2]{\frac{d #1}{d #2}}
\newcommand{\diag}{\mathop{\rm diag}}
\newcommand{\Po}{\mathop{\rm P}}
\newcommand{\Ucl}{U_\cl}

\newcommand{\Mfour}{M_4}
\newcommand{\Mfive}{M_5}
\newcommand{\Msix}{M_6}
\newcommand{\RMfour}{\bR\times M_4}
\newcommand{\SMfour}{S^1\times M_4}
\newcommand{\RMfourI}{\bR\times M_4\times[0,1]}
\newcommand{\SMfourI}{S^1\times M_4\times[0,1]}
\newcommand{\Mfourzero}{M_4^{(0)}}
\newcommand{\Mfourinfty}{M_4^{(\infty)}}
\newcommand{\Mfivezero}{M_5^{(0)}}
\newcommand{\Mfiveinfty}{M_5^{(\infty)}}
\newcommand{\Msixzero}{M_6^{(0)}}
\newcommand{\Msixinfty}{M_6^{(\infty)}}
\newcommand{\Dtwo}{D_2}

\newcommand{\Iz}{{\mathcal I}_1}
\newcommand{\Izp}{{\mathcal I}_2}
\newcommand{\LCS}{L_{\rm CS}}
\newcommand{\ag}{\xi}
\newcommand{\h}{U}
\newcommand{\Hrhotot}{H_\rho^{\rm tot}}
\newcommand{\dd}{\eta}
\newcommand{\Q}{K}
\newcommand{\MeV}{\text{MeV}}
\newcommand{\MKK}{M_{\rm KK}}
\newcommand{\gXt}{\wt{g}}


\begin{titlepage}

\title{
\hfill\parbox{4cm}
{\normalsize
KUNS-2103\\ YITP-07-62}\\
\vspace{1cm}
{\bf
Baryons and the Chern-Simons term\\
in holographic QCD with three flavors
}}

\author{
Hiroyuki {\sc Hata}$^{a}$\,\thanks{
{\tt hata@gauge.scphys.kyoto-u.ac.jp}}
\ and
Masaki {\sc Murata}$^{b}$\,\thanks{
{\tt masaki@yukawa.kyoto-u.ac.jp}}
\\[7mm]
$^{a}${\it
Department of Physics, Kyoto University, Kyoto 606-8502, Japan
}\\[3mm]
$^{b}${\it Yukawa Institute for Theoretical Physics, Kyoto University}\\
{\it Kyoto 606-8502, Japan}
}
\date{{\normalsize October 2007}}
\maketitle

\begin{abstract}
\normalsize
We study dynamical baryons in the holographic QCD model of Sakai and
Sugimoto in the case of three flavors and with special interest in the
construction of the Chern-Simons (CS) term. The baryon classical
solution in this model is given by the BPST instanton, and we carry
out the collective coordinate quantization of the solution.
The CS term should give rise to a first class constraint which selects
baryon states with right spins. However, the original CS term written
in terms of the CS 5-form does not work. We instead propose a new CS
term which is gauge invariant and is given as an integral over a six
dimensional space having as its boundary the original five dimensional
spacetime of the holographic model. Collective coordinate quantization
using our new CS term leads to correct baryon states and their mass
formula.

\end{abstract}

\thispagestyle{empty}
\end{titlepage}

\section{Introduction}
\label{Intro}

Among various approaches to the holographic dual of large $N_c$
QCD, the model proposed by Sakai and Sugimoto \cite{SaSu1,SaSu2} is
one of the most successful ones at present both theoretically and
phenomenologically. This model with $N_f$ massless quarks is
constructed using the brane configuration of $N_c$ D4-branes and $N_f$
D8-branes in type IIA superstring theory. They analyzed the effective
theory of D8-branes on the D4-brane background, which is a $U(N_f)$
Yang-Mills (YM) theory with Chern-Simons (CS) term on a curved
five-dimensional background. They found that this model has massless
pion as the Nambu-Goldstone boson of chiral symmetry breaking and
infinite number of massive (axial-)vector mesons. It well
reproduces various phenomenologically important parameters such as the
masses and the couplings of the mesons.
Moreover, when we truncate all the massive modes, then the effective
theory is found to be the Skyrme model \cite{Skyrme:1} with
Wess-Zumino-Witten (WZW) term \cite{Wess-Zumino,Witten:WZW}, which is
known as the effective theory of massless mesons.

The Sakai-Sugimoto model (SS-model) can also describe the baryon
degrees of freedom.
It has been argued that, in the AdS/CFT correspondence, a baryon
is identified as a D-brane wrapped around a sphere
\cite{Witten:1998xy}.
In the SS-model, this D-brane, which is a D4-brane wrapped on
$S^4$ in the color D4-brane background, is realized as a soliton in
the effective theory of D8-brane, namely, the five dimensional YM+CS
theory. Therefore, when we quantize the collective coordinates of the
instanton, the baryon spectra are expected to appear as in the case of
the Skyrme model \cite{ANW}.
In \cite{HSSY}, explicit construction of the baryon solution in
the YM+CS theory and its collective coordinate quantization were
carried out in the case of $N_f=2$ in the approximation of large
't\,Hooft coupling $\lambda\gg 1$ (see also \cite{HRYY1,HRYY2} for the
construction of the solution). The baryon solution at a fixed
time was found to be the BPST instanton solution \cite{BPST} with its
size of order $\lambda^{-1/2}$ determined by the energy balance:
the curved color D4-brane background tends to shrink the instanton
size, while the Coulomb self-energy from the CS term favors larger
instanton. Quantization of the collective coordinates including the
size of the instanton leads to the baryon spectra which agree fairly
well with experiments by taking a suitable Kaluza-Klein mass scale
$\MKK$ of the theory.

The purpose of this paper is to extend the study of \cite{HSSY} to the
case of three flavors, $N_f=3$. In fact, this is not a simple
problem.
First, all the quarks are massless in the original SS-model,
and we have to modify the model to give at least the strange quark a
mass. This is absolutely necessary for the comparison of the
model with experiments. Though there have appeared a number of
proposals to generate quark and meson masses in the SS-model
\cite{HHM,Evans,BSS,Dhar}, concrete calculation seems not easy at
present.
In this paper we focus on another problem in the $N_f=3$
SS-model, namely, the problem associated with the CS term.
As we mentioned above, the $U(1)$ part of the CS term plays an
important role in giving the instanton a non-vanishing size already in
the $N_f=2$ case. On the other hand, the $SU(N_f)$ part of the CS term
vanishes identically in the $N_f=2$ case, and $N_f=3$ is the first
nontrivial place where the non-abelian part of the CS term enters
the analysis of the theory.

To explain the problem of the CS term in the $N_f=3$ SS-model, let us
recall the role of the WZW term in the quantization of the collective
coordinate of the $SU(3)$ rotation of the baryon solution in the
$N_f=3$ Skyrme model \cite{Guadagnini,Mazur,Chemtob,Jain,Manohar}
(the WZW term vanishes identically in the $N_f=2$ case).
In this case, there arises a first class constraint
\begin{equation}
J_8=\frac{N_c}{2\sqrt{3}} \ ,
\label{eq:J_8-constraint}
\end{equation}
where $J_8$ is the eighth-component of the charge of $SU(3)_J$, whose
first three components $(J_1,J_2,J_3)$ constitute the $SU(2)$ of space
rotation, and the RHS, $N_c/(2\sqrt{3})$, is from the WZW term.
The constraint \eqref{eq:J_8-constraint} selects the correct baryon
states with spin $1/2$ for the flavor octet and those with spin $3/2$
for the decuplet from the $SU(3)_J$ octet and decuplet, respectively,
containing also other states with wrong spins.

In the SS-model, the CS term should play the role of the
WZW term in the Skyrme model (recall that the WZW term is reproduced
from the CS term in the low energy limit \cite{SaSu1}).
However, in collective coordinate quantization of the baryon solution
in the SS-model with $N_f=3$, the CS term originally proposed in
\cite{SaSu1,SaSu2} (given as \eqref{eq:SCS_SS} in sec.\
\ref{sec:baryonsolution}) vanishes identically as we will see
in sec.\ \ref{sec:LCC}.
This implies the absurd result that the constraint in the SS-model is
$J_8=0$ instead of \eqref{eq:J_8-constraint}.

To overcome this difficulty, we propose a new CS term for the
SS-model (see eq.\ \eqref{eq:newSCS}). Our new CS term is strictly
gauge invariant, in contrast to the original CS term of
\cite{SaSu1,SaSu2} which is not invariant under ``large'' gauge
transformations. However, for defining our CS term, we need a
fictitious sixth coordinate just as the WZW term needs the fifth
coordinate.
With our new CS term, we can carry out the collective coordinate
quantization of the baryon solution and get the desired constraint
\eqref{eq:J_8-constraint}. The two CS terms, \eqref{eq:SCS_SS} and
\eqref{eq:newSCS}, are naively the same if we use the relation
$\tr\cF^3=d\omega_5(\cA)$. The reason why the two CS terms lead to
different results is that the BPST instanton solution needs two
patches for describing it in the whole four-dimensional space
including both the origin and the infinity, and hence the space of
integration for \eqref{eq:SCS_SS} is not the only boundary of that for
\eqref{eq:newSCS} (see appendix \ref{app:another} for details).
In this paper, we introduce the sixth dimension for our CS term simply
by hand. It is interesting if this extra dimension has its origin in
ten dimensions of IIA superstring theory, though this seems not so
easy as we discuss in sec.\ \ref{sec:summary}.

This paper is organized as follows.
In sec.\ \ref{sec:baryonsolution}, we write down our model, five
dimensional $U(N_f)$ YM+CS theory in curved background, and obtain the
classical solution representing a baryon. We keep $N_f$ generic in
this section, and put $N_f=3$ in sec.\ \ref{sec:LCC} and later.
In sec.\ \ref{sec:LCC}, we introduce the collective coordinates into
the baryon solution and obtain their lagrangian for the case $N_f=3$.
There, we find that the original CS term does not work.
We also find that the WZW term obtained from this CS term in the low
energy limit cannot reproduce the constraint \eqref{eq:J_8-constraint}
either.
In sec.\ \ref{sec:NewCS}, we propose our new CS term and show that it
leads to the constraint \eqref{eq:J_8-constraint}.
Then, in sec.\ \ref{sec:CCQ}, we complete the collective coordinate
quantization using our new CS term and obtain the baryon mass formula.
We also make a brief comparison of this formula with experimental
data, though we have to introduce the strange quark mass for more
serious analyses.
The final section (sec.\ \ref{sec:summary}) is devoted to a summary
and discussions.
The appendices contain various technical details. In particular,
in appendices \ref{app:another} and \ref{app:WZWfromnewSCS},
we present details concerning our new CS term.

\section{SS-model with $\bm{N_f}$ flavors and the baryon solution}
\label{sec:baryonsolution}

In this section, we recapitulate the action of the SS-model with
$N_f$ flavors and obtain its classical solution representing a
baryon. Although in this paper we are eventually interested in the
case of three flavors, $N_f=3$, we keep $N_f$ generic in this
section.

\subsection{The action of the SS-model}
\label{setup}

We consider the effective theory of $N_f$ probe D8-branes in the
background of $N_c$ D4-branes \cite{SaSu1,SaSu2}.
Discarding the dependence on the $S^4$ around which the D8-branes are
wrapped, this effective theory is a $U(N_f)$ gauge theory in the
five dimensional subspace of the world volume of the D8-branes.
The $U(N_f)$ gauge field $\cA$, which is hermitian and corresponds to
the open string with both ends attached to the D8-branes, is given by
\begin{equation}
\cA = \cA_\mu dx^\mu + \cA_zdz \  ,
\end{equation}
where $\mu,\nu=0,1,2,3$ are four-dimensional Lorentz indices and $z$
is the coordinate of the fifth-dimension.
The action of the theory consists of the Yang-Mills (YM) part
$S_{\rm YM}$ and the Chern-Simons (CS) part $\SCS$,
\begin{equation}
S=S_{\rm YM}+S_{\rm CS}\ ,
\label{model}
\end{equation}
with
\begin{align}
S_{\rm YM}[\cA]&=-\kappa
\int d^4 x dz\,\tr\left[\,
\frac12\,h(z){\cF}_{\mu\nu}^2+k(z){\cF}_{\mu z}^2
\right]\ ,
\label{eq:SYM}
\\
S_{\rm CS}[\cA]&=\frac{N_c}{24\pi^2}
\int_{\Mfive}\omega_5^{U(N_f)}({\cA})\ ,
\label{eq:SCS_SS}
\end{align}
where $\cF=d\cA+i\cA^2$ is the field strength, and
$\omega_5^{U(N_f)}(\cA)$ is the CS 5-form defined by
\begin{equation}
\omega_5^{U(N_f)}({\cA})=\tr\left(
\cA \cF^2-\frac{i}{2}\cA^3\cF-\frac{1}{10}\cA^5
\right)\ .
\label{eq:omega_5}
\end{equation}
In $S_{\rm YM}$ \eqref{eq:SYM}, $\kappa$ is written by the 't\,Hooft
coupling $\lambda$ and the number of colors $N_c$ as
\begin{equation}
\kappa=a\lambda N_c\ ,
\quad\left(a=\frac{1}{216\pi^3}\right)\ ,
\label{kappa}
\end{equation}
and $h(z)$ and $k(z)$ are the warp factors given by
\begin{equation}
h(z)=(1+z^2)^{-1/3}\ ,\quad k(z)=1+z^2\ .
\label{hk}
\end{equation}
The space of integration in \eqref{eq:SCS_SS} (and also in
\eqref{eq:SYM}) is $\Mfive=\RMfour$ with $\bR$ for the time $t$ and
$\Mfour$ for $(\bm{x},z)$.
Here, we adopt the original CS term \eqref{eq:SCS_SS} of
\cite{SaSu1,SaSu2}. Although we need a refinement on the definition of
the CS term for the proper quantization around the baryon solution,
the present one \eqref{eq:SCS_SS} is sufficient for obtaining
classical solutions since the equations of motion (EOM) are not
affected by the redefinition of the CS term.

Let us decompose the $U(N_f)$ gauge field $\cA$
into the $SU(N_f)$ part $A$ and the $U(1)$ part $\Ah$ as
\begin{equation}
\cA=A+\frac{1}{\sqrt{2N_f}}\,\Ah
=A^at_a+\frac{1}{\sqrt{2N_f}}\,\Ah\ ,
\end{equation}
where $t_a$ ($a=1,2,\cdots,N_f^2-1$) are the hermitian generators
of $SU(N_f)$ normalized as
\begin{equation}
\tr(t_at_b)=\frac12\delta_{ab}\ .
\end{equation}
Using $A$ and $\Ah$, the actions $\SYM$ and $\SCS$ read
\begin{align}
S_{\rm YM}&=-\kappa\int d^4xdz\tr\left[\frac12 h(z)F_{\mu\nu}^2
+ k(z)F_{\mu z}^2\right]
-\frac12\kappa\int d^4xdz\left[
\frac12h(z)\Fh_{\mu\nu}^2+ k(z)\Fh_{\mu z}^2\right]\ ,
\\[3mm]
S_{\rm{CS}}&=\frac{N_c}{24\pi^2}\int\biggl[\,
\omega_5^{SU(N_f)}(A)
+\frac{1}{\sqrt{2N_f}}\!\left(3\Ah\tr F^2 +\frac12\Ah\Fh^2
\right)
+\frac{1}{\sqrt{2N_f}}\,
d\!\left(\Ah\tr\!\left(2FA-\frac{i}{2}A^3\right)\right)
\biggr] \notag\\
&= \frac{N_c}{24\pi^2}\int\omega_5^{SU(N_f)}(A)
+ \frac{N_c}{24\pi^2}\sqrt{\frac2{N_f}}\epsilon_{MNPQ}
\int d^4xdz\biggl[
\frac38\Ah_0\tr(F_{MN}F_{PQ}) \notag\\
&\qquad- \frac32\Ah_M\tr(\partial_0A_NF_{PQ})
+ \frac34\Fh_{MN}\tr(A_0F_{PQ})
+ \frac1{16}\Ah_0\Fh_{MN}\Fh_{PQ} \notag\\
&\qquad - \frac14\Ah_M\Fh_{MN}\Fh_{PQ}
+ \text{(total derivatives)}\biggl]
\ ,
\label{CS}
\end{align}
with $M,N=1,2,3,z$ and $\epsilon_{123z}=+1$.
The genuine non-abelian part $\omega_5^{SU(N_f)}(A)$ is missing in the
$N_f=2$ case.

\subsection{Classical solution representing a baryon}
\label{solution}

In this subsection, we obtain the classical solution of the SS-model
representing a baryon in the $1/\lambda$ expansion by assuming that
the 't\,Hooft coupling $\lambda$ is large enough. The number of
flavors $N_f$ is kept generic, not restricted to the $N_f=3$ case.
Our solution is an extension of the $N_f=2$ solution of \cite{HSSY} to
a generic $N_f$,\footnote{
The baryon solution in the $N_f=2$ case is also analyzed in
\cite{HRYY1}. Moreover, the energy \eqref{smass} and the size
\eqref{rhomin} of the solution for a generic $N_f$ have already been
obtained in \cite{HRYY2}.
}
and it is essentially the embedding of the $SU(2)$
BPST instanton solution to $SU(N_f)$.
A nontrivial point in the $N_f\ge 3$ case is
the appearance of the time component $A_0$ of the $SU(N_f)$ part of
the gauge field, which is absent in the $N_f=2$ case.

In order to carry out a systematic $1/\lambda$ expansion, we
follow ref.\ \cite{HSSY} to rescale the coordinates $x^M=(\bm{x},z)$
and the gauge field $\cA$ as
\begin{align}
&x^M\to\lambda^{+1/2}x^M \ ,\quad x^0\to x^0 \ ,
\nn\\
&\cA_0\to \cA_0 \ ,\quad
\cA_M\to \lambda^{-1/2}\cA_M \ ,
\nn\\
&\cF_{MN}\to \lambda^{-1}\cF_{MN} \ ,\quad
\cF_{0M}\to \lambda^{-1/2}\cF_{0M} \ .
\label{rescaling}
\end{align}
Note that $S_{\rm CS}$ is invariant under this rescaling, while
$S_{\rm YM}$ is expanded as
\begin{align}
S_{{\rm YM}}=&-aN_c\int d^4 x dz \,\tr\left[\,
\frac{\lambda}{2}\,F_{MN}^2+\left(
-\frac{z^2}{6} F_{ij}^2+z^2 F_{iz}^2- F_{0M}^2
\right)+\cO(\lambda^{-1})
\right]
\nn\\
&-\frac{aN_c}{2}\int d^4 x dz \,
\left[\,\frac{\lambda}{2}
\, \Fh_{MN}^2+
\left(
-\frac{z^2}{6} \Fh_{ij}^2+z^2 \Fh_{iz}^2-\Fh_{0M}^2
\right)+\cO(\lambda^{-1})\right]\ ,
\label{DBI}
\end{align}
with $i,j=1,2,3$.
Here, we have used \eqref{kappa} for $\kappa$.
{}From this action, the EOM reads as follows:
\begin{align}
&D_M F_{0M}+\frac{1}{64\pi^2 a}\sqrt{\frac2{N_f}}
\epsilon_{MNPQ}\Fh_{MN}F_{PQ} \nn\\
&\qquad\qquad\qquad+ \frac1{64\pi^2a}\epsilon_{MNPQ}\left\{
F_{MN}F_{PQ}-\frac1{N_f}\tr(F_{MN}F_{PQ})\right\}
+\cO(\lambda^{-1})=0 \ ,
\label{Gausslaw}\\
&D_NF_{MN}+\cO(\lambda^{-1})=0 \ ,
\label{eom:su}\\
&\p_M\Fh_{0M}+\frac{1}{64\pi^2 a}\sqrt{\frac2{N_f}}
\epsilon_{MNPQ}\left\{
\tr(F_{MN}F_{PQ})+\frac12\Fh_{MN}\Fh_{PQ}
\right\}+\cO(\lambda^{-1})=0 \ ,
\label{Gauss:u1}\\
&\p_{N}\Fh_{MN}+\cO(\lambda^{-1})=0 \ ,
\label{eom:u1}
\end{align}
where \eqref{Gausslaw} and \eqref{eom:su} are the EOM
for the $SU(N_f)$ part, while \eqref{Gauss:u1} and \eqref{eom:u1} are
for the $U(1)$ part.

Let us obtain the static soliton solution of the EOM
\eqref{Gausslaw}--\eqref{eom:u1} corresponding to a baryon.
In this paper, we want to construct the solution so that
its energy is correctly obtained to next to the leading order in the
$1/\lambda$ expansion.
First, let us solve \eqref{eom:su}.
For the purpose of the present paper it is sufficient to consider the
leading part $D_NF_{MN}=0$, and the solution carrying a unit baryon
number is given by the embedding of the $SU(2)$ BPST instanton
solution \cite{BPST} in the flat four-dimensional space to $SU(N_f)$:
\begin{equation}
A_M^\cl(x)=-if(\xi)\,g(x)\p_M g(x)^{-1} \ ,
\label{eq:BPST}
\end{equation}
where $f(\xi)$ and $g(x)$ are given by\footnote{
We have chosen $g^{SU(2)}(x)$ \eqref{eq:g} as the hermitian conjugate
of $g(x)$ in ref.\ \cite{HSSY} so as to make the corresponding
$A^\cl_M$ \eqref{eq:BPST} carry a unit baryon number $N_B=+1$ (see
\eqref{eq:N_B=1}).
}
\begin{align}
f(\xi)&=\frac{\xi^2}{\xi^2+\rho^2} \ ,\quad
\xi=\sqrt{(x^M-X^M)^2} \ ,
\label{eq:f}
\\
g(x)&
=\begin{pmatrix}
g^{SU(2)}(x) & 0 \\ 0 & \bm{1}_{N_f-2}
\end{pmatrix}
, \quad
g^{SU(2)}(x)=
\frac1{\xi}\left((z-Z)\bm{1}_2 + i(x^i-X^i)\tau_i\right)\ .
\label{eq:g}
\end{align}
Here, $\bm{1}_{N}$ denotes the $N\times N$ identity matrix, and
$\tau_i$ ($i=1,2,3$) are the Pauli matrices.
The constants $X^M=(\bm{X},Z)$ and $\rho$ represent the position and
the size of the instanton, respectively.
Notice that these constants are also rescaled as in \eqref{rescaling}.
The field strengths of this solution are given by
\begin{equation}
F^\cl_{ij}=\frac{4\rho^2}{(\xi^2+\rho^2)^2}\epsilon_{ijk}t_k \ ,
\quad
F^\cl_{iz}=\frac{4\rho^2}{(\xi^2+\rho^2)^2}\,t_i \ ,
\label{Fcl}
\end{equation}
where $t_i$ is the $SU(N_f)$ embedding of $\tau_i$,
$t_i=\frac12\!\sPmatrix{\ds \tau_i & \ds 0\\ \ds 0 & \ds 0}$.

Next, the solutions to the $U(1)$ part of EOM, \eqref{eom:u1} and
\eqref{Gauss:u1}, are the same as in the $SU(2)$ case \cite{HSSY}.
We have
\begin{equation}
\Ah^\cl_M=0\ ,
\label{hatAM}
\end{equation}
and
\begin{equation}
\Ah^\cl_0 = \sqrt{\frac2{N_f}}\frac{1}{8\pi^2a}\frac1{\xi^2}
\left(1-\frac{\rho^4}{(\xi^2+\rho^2)^2}\right) .
\label{A0hat}
\end{equation}
The present $\Ah^\cl_0$ has been chosen to be regular at the origin
$\xi=0$ and vanish at the infinity $\xi\to\infty$.

Finally, let us solve \eqref{Gausslaw} to obtain $A_0$.
In the $N_f=2$ case, the third term of \eqref{Gausslaw} is missing,
and the solution vanishing at $\xi=\infty$ is simply given by
$A_0=0$. For a generic $N_f$, substituting \eqref{eq:BPST} and
\eqref{hatAM} into \eqref{Gausslaw}, we have
\begin{equation}
D_M^2A_0 + \frac{3}{2\pi^2a}\frac{\rho^4}{(\xi^2+\rho^2)^2}
\left(\cP_2-\frac2{N_f}\bm{1}_{N_f}\right) = 0 \ ,
\label{GausslawBPST}
\end{equation}
where the matrix $\cP_2$ is $\cP_2=\diag(1,1,0,\cdots,0)$.
Eq.\ \eqref{GausslawBPST} leads to the following nontrivial regular
solution which commutes with $A^\cl_M$ \eqref{eq:BPST}, vanishes at
the infinity, and has the same $\xi$-dependence as that of
\eqref{A0hat}:
\begin{equation}
A^\cl_0 = \frac1{16\pi^2a}\frac1{\xi^2}\left(
1-\frac{\rho^4}{(\xi^2+\rho^2)^2}\right)
\left(\cP_2-\frac2{N_f}\bm{1}_{N_f}\right) \ .
\label{A0}
\end{equation}

The mass $M$ of our static soliton solution is obtained by using the
relation $S=-\int dtM$.
Substituting the above solution into \eqref{DBI} and \eqref{CS}, we
get
\begin{align}
M&=\kappa\int\! d^3xdz\tr
\left[\frac12(F^\cl_{MN})^2
-\lambda^{-1}\!\left(\frac{z^2}{6}(F^\cl_{ij})^2
+z^2(F^\cl_{iz})^2-(F^\cl_{0M})^2\right)\right]
-\frac{\kappa}2\lambda^{-1}\!\int\! d^3xdz(\Fh^\cl_{0M})^2
\nn\\
&\qquad
-\frac{\kappa}{24\pi^2a}\lambda^{-1}\epsilon_{MNPQ}\int\!d^3xdz\!
\left[\sqrt{\frac2{N_f}}\frac38\Ah^\cl_0\tr(F^\cl_{MN}F^\cl_{PQ})
+\frac34\tr(A^\cl_0F^\cl_{MN}F^\cl_{PQ})
\right]
+\cO(\lambda^{-1})\nn\\
&=8\pi^2\kappa\left[
1+\lambda^{-1}\left(\frac{\rho^2}{6}
+\frac{1}{320\pi^4a^2}
\frac{1}{\rho^2}+\frac{Z^2}{3}\right)
+\cO(\lambda^{-2})\right] \ .
\label{smass}
\end{align}
The contributions from the two terms, $\tr(F^\cl_{0M})^2$ and
$\tr(A^\cl_0F^\cl_{MN}F^\cl_{PQ})$, are absent in the $N_f=2$ case
\cite{HSSY}.
It is interesting that the mass formula \eqref{smass} is nonetheless
independent of the number of flavors $N_f$.
The values of $\rho$ and $Z$ for the stable solution is determined
by minimizing $M$:\footnote{
This is equivalent to solving the sub-leading part of the EOM
\eqref{eom:su} and \eqref{eom:u1} projected on to the subspace of
deformations of the solution in the $\rho$ and $Z$ directions.
}
\begin{equation}
\rho^2=\frac{1}{8\pi^2 a}\sqrt{\frac{6}{5}} \ ,
\qquad
Z=0 \ .
\label{rhomin}
\end{equation}
Note that the size of the instanton is also independent of $N_f$.
If we express this in terms of the original variable (see
\eqref{rescaling}), $\rho^2$ is rescaled as
$\rho^2\to\lambda^{+1}\rho^2$.
This fact means that the size of our solution is of order
$\lambda^{-1/2}$.
Inserting \eqref{rhomin} into \eqref{smass}, the mass of the soliton
is given by
\begin{equation}
M= 8\pi^2\kappa
+\sqrt{\frac{2}{15}}N_c \ .
\label{Mcl}
\end{equation}
The very small size of order $\lambda^{-1/2}$ of the baryon solution
implies that the higher order derivative terms in the D-brane
effective action, which have been neglected in \eqref{eq:SYM}, might
have important contributions as mentioned in \cite{HSSY}.
However, we leave this issue for future study and continue analysis
based on the YM action \eqref{eq:SYM} in the rest of this paper.

\section{Necessity of modifying the CS term}
\label{sec:LCC}

Having constructed the baryon classical solution in sec.\
\ref{sec:baryonsolution}, our next task is to carry out the
quantization of the collective coordinates of the solution.
However, as we mentioned in the Introduction, there arise a problem
that, in the $N_f=3$ case, the constraint \eqref{eq:J_8-constraint}
necessary for selecting the baryon states with correct spins cannot
be obtained from the CS term \eqref{eq:SCS_SS} of \cite{SaSu1,SaSu2}.

In this section, we first introduce the collective coordinates into
our baryon classical solution (sec.\ \ref{subsec:IntroCC}),
obtain the lagrangian of collective coordinates
(sec.\ \ref{subsec:LCC}), and then explain how the CS term
\eqref{eq:SCS_SS} fails to give the constraint
\eqref{eq:J_8-constraint} (sec.\ \ref{subsec:CStermfails}).
We also show that the WZW term obtained as the low energy limit of the
CS term \eqref{eq:SCS_SS} cannot reproduce the constraint
\eqref{eq:J_8-constraint} either (sec.\ \ref{subsec:WZWtermfails}).
In the rest of this paper, we restrict ourselves to the three flavor
case, $N_f=3$.

\subsection{Introducing the collective coordinates}
\label{subsec:IntroCC}

We take the following moduli of the classical solution as the
collective coordinates for quantization:
\vspace{-1mm}
\begin{itemize}
	\item $SU(3)$ orientation $\A\in SU(3)$
	\item Size of the instanton $\rho$
	\item Position of the instanton $X^M=(\bm{X},Z)$
\end{itemize}
\vspace{-1mm}
Namely, we analyze the quantum mechanical system consisting of the
above three kinds of moduli promoted to time-dependent variables
$(\A(t), X^M(t), \rho(t))$.
Note that $\rho$ and $Z$ are not genuine moduli as seen from the fact
that the mass \eqref{smass} of the solution depends on them.
However, as in the $N_f=2$ case, the masses of the modes $\rho$ and
$Z$ are much lighter than other massive modes for large $\lambda$.
Therefore, we regard $\rho$ and $Z$ as the collective coordinates as
well as $\A$ and $\bm{X}$.

In order to derive the lagrangian of these collective modes, we
approximate the slowly moving soliton by the static solution
of the last section with $X^\alpha=(X^M,\rho)$ and the $SU(3)$
orientation $\A$ made time-dependent.
Thus, the $SU(3)$ gauge field is assumed to be of the form\footnote{
Here, we adopt a different way of introducing the collective
coordinate of $SU(3)$ rotation from that of ref.\ \cite{HSSY}.
The gauge field \eqref{def:coll} in this paper and the corresponding
one (4.2) in \cite{HSSY} (extended to the
$N_f=3$ case) are related through the gauge transformation by
$Y(t,x)$ defined by $-iY^{-1}\dot{Y}=\Delta A_0$.
The variable $V$ in ref.\ \cite{HSSY} and $W$ in this paper are
related by $V(t,x)=Y(t,x)\A(t)$.
}
\begin{align}
A_M(t,x)&=
\A(t)A^\cl_M(x;X^\alpha(t))\A(t)^{-1} \ ,
\nn\\
A_0(t,x)&= \A(t)A^\cl_0(x;X^\alpha(t))\A(t)^{-1}
+\Delta A_0(t,x) \ ,
\label{def:coll}
\end{align}
where $A^\cl_M(x;X^\alpha(t))$ is the BPST instanton solution
\eqref{eq:BPST} with time-dependent $X^\alpha$.
The $U(1)$ part of the gauge field, $\Ah_M(x,t)$ and $\Ah_0(x,t)$, are
given simply by \eqref{hatAM} and \eqref{A0hat}, respectively, with
$X^\alpha$ made time-dependent:
\begin{equation}
\Ah_M(x,t)=0 \ ,
\qquad
\Ah_0(x,t)=\Ah^\cl_0\bigl(x;X^\alpha(t)\bigr) \ .
\label{eq:Ah(x,t)}
\end{equation}

The extra term $\Delta A_0(x,t)$ in \eqref{def:coll} for $A_0$ is
introduced so that the EOM of $A_0$, namely, the Gauss law constraint
\eqref{Gausslaw}, is satisfied for the present gauge field with
time-dependent moduli.\footnote{
The general principle of introducing the time-dependent collective
coordinates into a classical solution is that the EOM of the
collective coordinates ensure the field theory EOM. In gauge theories,
this requirement is automatically satisfied except for the EOM of
$A_0$. For $A_0$ we have to add an extra term to ensure its EOM by
hand. We would like to thank S.\ Sugimoto, T.\ Sakai and S.\ Yamato
for discussions on this matter.
}
Let us see how $\Delta A_0$ is determined.
For $A(x,t)$ of \eqref{def:coll}, we find that
\begin{align}
F_{MN}&=\A(t)F_{MN}^\cl\A(t)^{-1} \ ,
\label{eq:F_MN}
\\
F_{0M}&=\A(t)\left(
\dot X^\alpha\frac{\p}{\p X^\alpha}A_{M}^\cl
-D_M^\cl\Phi- D_M^\cl A_0^\cl\right)\A(t)^{-1} \ ,
\label{eq:F_0M}
\end{align}
where $\Phi(t,x)$ is defined by
\begin{equation}
\Phi(t,x)=\A(t)^{-1}\Delta A_0\,\A(t) - i\A(t)^{-1}\dot{\A}(t) \ .
\label{def:Phi}
\end{equation}
Then, \eqref{Gausslaw} implies
\begin{equation}
D_M^{\rm cl}\left(\dot X^N
\frac{\p}{\p X^N} A_M^{\rm cl}
+\dot\rho \frac{\p}{\p \rho} A_M^{\rm cl}
-D_M^{\rm cl}\Phi\right)=0\ ,
\label{Gausslaw2}
\end{equation}
and the problem of determining $\Delta A_0$ has been reduced to that
of solving \eqref{Gausslaw2} for $\Phi$.

The solution to \eqref{Gausslaw2} is given as the sum of three terms,
$\Phi=\Phi_X + \Phi_\rho +\Phi_{SU(3)}$, each of which depends on
the time derivative of the corresponding collective coordinate.
The determination of the solution $\Phi$ is explained in appendix A
of ref.\ \cite{HSSY} in the case of $N_f=2$.
In the present $N_f=3$ case, $\Phi_X$ and $\Phi_\rho$ remain the same
as in the $N_f=2$ case,
$\Phi_X=-\dot{X}^N A^\cl_N$ and $\Phi_\rho=0$,
and we have only to solve $D_M^\cl D_M^\cl\Phi_{SU(3)}=0$.
Derivation of $\Phi_{SU(3)}$ is explained in appendix
\ref{derive:Phi}, and we find that $\Phi$ in the $N_f=3$ case is
\begin{equation}
\Phi(t,x)=-\dot X^N(t)A_N^{\rm cl}(x;X^\alpha(t))
+ \chi^a(t)\Phi_a(x;X^\alpha(t)) \ ,
\label{sol:Phi}
\end{equation}
where $\Phi_a(x;X^\alpha(t))$ ($a=1,\ldots,8$) are given by
\eqref{sol:Phia} in terms of $u^a(\xi)$ of \eqref{eq:u^a}, and
$\chi^a(t)$ are arbitrary.
In order to relate $\chi^a(t)$ to $\A(t)$, we impose the
condition,\footnote{
The condition \eqref{eq:DeltaA_0to0} with $z\to +\infty$ only may look
strange. In fact, $\Delta A_0(t,x)$ of \eqref{def:coll}
and hence $A_0(t,x)$ itself does not tend to zero in the other limit
$z\to -\infty$ since $g(x)\to\diag(-1,-1,1)\ne\bm{1}_3$ in this
limit. Eq.\ \eqref{eq:DeltaA_0to0} should be regarded as a consequence
of the condition $\ol{A}_0(t,x)\to 0$  ($\xi\to\infty$) requesting
that the gauge field $\ol{A}_0$ in the patch containing the infinity
$\xi=\infty$ be regular there. See appendix \ref{subapp:M_5}.
}
\begin{equation}
\Delta A_0(t,x) \rightarrow 0\quad\text{as}\quad
z\to +\infty .
\label{eq:DeltaA_0to0}
\end{equation}
Then, since we have $\Phi_a(x)\to t_a$ and $A^\cl_M(x)\to 0$ as
$z\to +\infty$, we obtain
\begin{equation}
\chi^a(t)=-2i\tr\bigl(t_a\A(t)^{-1}\dot{\A}(t)\bigr)\ .
\label{chi-A}
\end{equation}

Summarizing, we find that $F_{0M}$ is given by
\begin{equation}
F_{0M}=\A(t)\left(
\dot X^N F^\cl_{MN} +\dot{\rho}\frac{\p}{\p\rho}A^\cl_M
-\chi^a D_M^\cl\Phi_a-D_M^\cl A^\cl_0\right)\A(t)^{-1}\ ,
\label{eq:F_0M_summary}
\end{equation}
where we have used
$(\p/\p X^N)A_M^{\rm cl}=-\p_N A_M^{\rm cl}$.
The $SU(3)$ part of the gauge field 1-form $A(t,x)$ \eqref{def:coll}
is concisely expressed as
\begin{equation}
A(t,x) =\bigl(A^\cl(x;X^\alpha(t))+\Phi(t,x)dt\bigr)^{\A(t)}\ ,
\label{coll:cpt}
\end{equation}
where $A^V$ is the gauge transform of $A$ by $V(t,x)\in SU(3)$:
\begin{equation}
A^V=V(A-id)V^{-1}\ .
\label{eq:A^G}
\end{equation}
Since the $U(1)$ part $\Ah(x,t)$ is simply given by
\eqref{eq:Ah(x,t)}, the formula \eqref{coll:cpt} is extended to the
whole $\cA=A+\Ah$ as
\begin{equation}
\cA(t,x) =\bigl(\cA^\cl(x;X^\alpha(t))+\Phi(t,x)dt\bigr)^{\A(t)}\ .
\label{eq:cptU(3)}
\end{equation}

\subsection{Lagrangian of the collective coordinates}
\label{subsec:LCC}

The lagrangian $L$ of the collective coordinates
$X^\alpha(t)=\bigl(\bm{X}(t),Z(t),\rho(t)\bigr)$ and $\A(t)$ is
obtained as $S_{\rm YM}+\SCS=\int\! dt\,L$
by substituting \eqref{eq:F_MN} and \eqref{eq:F_0M_summary}
into $S_{\rm YM}$ \eqref{DBI}:\footnote{
In obtaining the last expression of \eqref{eq:L=-M+int},
we have carried out the integration-by-parts for the term
$\tr(\cO_M D^\cl_M A^\cl_0)$ with
$\cO_M=\dot{X}^\alpha(\p/\p X^\alpha)A^\cl_M-D^\cl_M\Phi$
to change it into $-\tr(A^\cl_0D^\cl_M\cO_M)$, which vanishes due to
\eqref{Gausslaw2}. The surface term can be dropped since we have
$A^\cl_0\sim 1/\xi^2$ and $\cO_M\sim 1/\xi^3$ as $\xi\to\infty$.
}
\begin{align}
L&=-M +a N_c\int\! d^3x dz\tr\!\left(
F_{0M}^2-(F^\cl_{0M})^2\right)
+\LCS
\nn\\
&=-M +a N_c\int\! d^3x dz\tr\!\left(
\dot X^N F^\cl_{MN} +\dot{\rho}\frac{\p}{\p\rho}A^\cl_M
-\chi^a D_M^\cl\Phi_a\right)^2
+\LCS \ ,
\label{eq:L=-M+int}
\end{align}
where $\LCS$ is defined by
\begin{equation}
\SCS[\cA]-\SCS[\cA^\cl]=\int\! dt\,\LCS \ .
\label{eq:defLCS}
\end{equation}
Performing the integrations over $(\bm{x},z)$, we get
\begin{equation}
L=-M_0+\frac{m_X}{2}\dot{\bm{X}}^2
+L_Z +L_\rho + L_{\rho\A} +\LCS\ ,
\label{eq:L=sumL}
\end{equation}
where $L_Z$, $L_\rho$ and $L_{\rho\A}$ are given by
\begin{align}
L_Z&=\frac{m_Z}{2}\left(\dot{Z}^2-\omega_Z^2 Z^2\right) ,
\\
L_\rho&=\frac{m_\rho}{2}\left(
\dot{\rho}^2-\omega_\rho^2\rho^2\right)
-\frac{\Q}{m_\rho \rho^2} \ ,
\\
L_{\rho\A}&=m_\rho\,\rho^2\left(
\frac18\sum_{a=1}^3(\chi^a)^2+\frac{1}{16}\sum_{a=4}^7(\chi^a)^2
\right) ,
\nn\\
&=2\,\Iz(\rho)\sum_{a=1}^3\left[
\tr\bigl(-i\A^{-1}\dot{\A}\,t_a\bigr)\right]^2
+2\,\Izp(\rho)\sum_{a=4}^7\left[
\tr\bigl(-i\A^{-1}\dot{\A}\,t_a\bigr)\right]^2 ,
\end{align}
with the various quantities defined as follows:
\begin{align}
M_0 &= 8\pi^2\kappa,\\
m_X &= m_Z =\frac{m_\rho}2 = 8\pi^2\kappa\lambda^{-1}=8\pi^2 aN_c, \\
\omega_Z^2&=\frac23,\quad \omega_\rho^2=\frac16, \\
\Q &= \frac{N_c m_\rho}{40\pi^2a} = \frac25 N_c^2 , \\
\Iz(\rho)&=\frac14 m_\rho\rho^2 ,\quad
\Izp(\rho)=\frac18 m_\rho\rho^2 \ .
\end{align}
The expressions of $M_0$, $m_{X,Z,\rho}$, $\omega_{Z,\rho}^2$ and
$Q\equiv \Q/m_\rho$ are the same as in the $SU(2)$ case \cite{HSSY}.
The ratio of the moments of inertia, $\Izp(\rho)/\Iz(\rho)=1/2$, is
due to the powers $1$ and $1/2$ of $f(\xi)$ in $u^a(\xi)$
\eqref{eq:u^a} for $a=1,2,3$ and $a=4,\cdots,7$, respectively.

\subsection{The CS term (\ref{eq:SCS_SS})}
\label{subsec:CStermfails}

Let us evaluate the CS term \eqref{eq:SCS_SS} for the configuration
\eqref{eq:cptU(3)} to see the dependence on the collective coordinates
$\A(t)$ and $X^\alpha(t)$.
Using the formulas of $\omega_5$ which are summarized in
appendix \ref{app:formulas_omega_5}, we get
(the superscript $U(3)$ on $\omega_5$ will be omitted for simplicity),
\begin{align}
\omega_5(\cA)&=\omega_5\bigl((\cA^\cl+\Phi dt)^\A\bigr)
\nn\\
&=\omega_5(\cA^\cl+\Phi dt)+\frac{1}{10}\tr(-i\A^{-1}\dot\A dt)^5
+d\alpha_4(-i\A^{-1}\dot{\A}dt,\cA^\cl+\Phi dt)
\nn\\
&=\omega_5(\cA^\cl)+3\tr\bigl(\Phi dt\,(\cF^\cl)^2\bigr)
+d\beta\bigl(\Phi dt,\cA^\cl\bigr)
+d\alpha_4\bigl(-i\A^{-1}\dot{\A} dt,\cA^\cl\bigr)
\nn\\
&=\omega_5(\cA^\cl)+3\tr\bigl(\Phi dt\,(F^\cl)^2\bigr)
+d\beta\bigl(\Phi dt,A^\cl\bigr)
+d\alpha_4\bigl(-i\A^{-1}\dot{\A} dt,A^\cl\bigr) \ ,
\label{eq:eval_omega_5(cA)}
\end{align}
where $\beta$ and $\alpha_4$ are given in \eqref{eq:beta} and
\eqref{eq:alpha_4}.
In obtaining the last expression, we have used that
$\Ah^\cl_M(x;X^\alpha(t))=\Fh^\cl_{MN}(x;X^\alpha(t))=0$.

As we mentioned in the Introduction, the dependences of the CS term
\eqref{eq:SCS_SS} on the collective coordinates, in particular, on
$\A(t)$ cancel out among the last three terms of
\eqref{eq:eval_omega_5(cA)}. This is seen as follows.
First, note that, in the term $3\tr\bigl(\Phi dt\,(F^\cl)^2\bigr)$,
we have $\bigl(F^\cl\bigr)^2=\frac12\cP_2\tr\bigl(F^\cl\bigr)^2$
and
\begin{equation}
\tr\bigl(\Phi\cP_2\bigr)=\frac{1}{\sqrt{3}}\chi^8(t)\ ,
\end{equation}
where we have used \eqref{sol:Phi}, \eqref{sol:Phia} and
\begin{equation}
\cP_2 =
\begin{pmatrix}
1&0&0 \\ 0&1&0 \\ 0&0&0
\end{pmatrix}
= \frac2{\sqrt3}t_8 + \frac23\bm{1}_3,\quad
t_8 = \frac1{2\sqrt3}
\begin{pmatrix}
1&0&0 \\ 0&1&0 \\ 0&0&-2
\end{pmatrix}.
\label{eq:cP_2}
\end{equation}
Therefore, we obtain
\begin{equation}
\frac{N_c}{24\pi^2}\int_{\Mfive=\RMfour}
\!\! 3\tr\bigl(\Phi dt\,(F^\cl)^2\bigr)
=\frac{N_c}{24\pi^2}\frac{\sqrt{3}}{2}
\int\! dt\,\chi^8(t)\int_{\Mfour}\! \tr(F^\cl)^2
=\frac{N_c}{2\sqrt{3}}\int\! dt\,\chi^8(t) ,
\label{eq:intPhiF^2}
\end{equation}
where we have used that our classical solution has a unit baryon
number (=instanton number):
\begin{equation}
N_B=\frac{1}{8\pi^2}\int_{\Mfour}\tr(F^\cl)^2=1 \ .
\label{eq:N_B=1}
\end{equation}
Evaluation of $\int d\beta$ and $\int d\alpha_4$ are similar
and easier. We have, using
$(A^\cl)^3\to (-igdg^{-1})^3\propto\cP_2$ and
$F^\cl(x)\sim 1/\xi^4$ as $\xi\to\infty$,
\begin{align}
&\frac{N_c}{24\pi^2}\int_{\RMfour}d\beta\bigl(\Phi dt,A^\cl\bigr)
=\frac{N_c}{24\pi^2}
\int_{\RMfour}d\alpha_4\bigl(-i\A^{-1}\dot{\A} dt,A^\cl\bigr)
\nn\\
&=\frac{N_c}{24\pi^2}\frac{i}{4\sqrt{3}}
\int\! dt\,\chi^8(t)\int_{\p\Mfour}\tr\bigl(-igdg^{-1}\bigr)^3
=-\frac{N_c}{4\sqrt{3}}\int\! dt\,\chi^8(t) ,
\label{eq:intdbeta=intdalpha}
\end{align}
where we have used another expression of \eqref{eq:N_B=1}:
\begin{equation}
\frac{-i}{24\pi^2}\int_{S^3}\bigl(-igdg^{-1}\bigr)^3
=\frac12 \cP_2 \ .
\label{eq:int(-igdg^-1)^3=}
\end{equation}

{}From \eqref{eq:intPhiF^2} and \eqref{eq:intdbeta=intdalpha}, we find
that the sum of the contributions of the three terms in
\eqref{eq:eval_omega_5(cA)} cancels out as announced:\footnote{
The CS term \eqref{eq:SCS_SS} becomes more involved if we adopt the
way of introducing the collective coordinate of $SU(3)$ rotation by
the variable $V(t,x)$ given in ref.\ \cite{HSSY}.
In this case, we can show that the terms linear in $\chi^a(t)$ are
missing from \eqref{eq:SCS_SS}.
}
\begin{equation}
\SCS[\cA]=\SCS[\cA^\cl] \ .
\label{eq:intomega5(cA)=intomega5(cA^cl)}
\end{equation}
Namely, $\LCS$ \eqref{eq:defLCS} vanishes:
\begin{equation}
\LCS=0 \ .
\label{eq:LCS=0}
\end{equation}

\subsection{WZW term}
\label{subsec:WZWtermfails}

In ref.\ \cite{SaSu1}, they showed that the Skyrme action including
the WZW term can be correctly reproduced as the low energy limit of
the action \eqref{model} of holographic QCD.
In particular, the WZW term comes from the CS term of
\eqref{eq:SCS_SS} and is given by
\begin{equation}
S_{\rm WZW}=\frac{N_c}{240\pi^2}\int_{\RMfour}\tr L^5 \ ,
\label{eq:SWZW_SS}
\end{equation}
where the left-current 1-form $L$ is defined by
\begin{equation}
L=-i\,U(t,\bm{x},z)d U(t,\bm{x},z)^{-1} \ ,
\label{eq:L=-iUdU^-1}
\end{equation}
with
\begin{equation}
U(t,\bm{x},z)=\Po\exp\left(
i\int_{-\infty}^z\! dz'\cA_z(t,\bm{x},z')\right) .
\label{eq:U=PexpintAz}
\end{equation}
In this WZW term, the coordinate $z$ plays the role of the fifth
dimension with $z=\infty$ corresponding to the real four dimensional
space-time $(t,\bm{x})$.

In this subsection, we will show that this WZW term \eqref{eq:SWZW_SS}
cannot reproduce the desired constraint \eqref{eq:J_8-constraint}
either. This is, of course, consistent with the result of the last
subsection.
Inserting \eqref{def:coll} into \eqref{eq:U=PexpintAz}, we have
\begin{equation}
U(t,\bm{x},z)=\A(t)\Ucl(\bm{x},z)\A(t)^{-1} \ ,
\label{eq:U=AUclA^-1}
\end{equation}
where $\Ucl$ is given, using
$\cA^\cl_z(x)=\bm{x}\cdot\bm{\tau}/(\xi^2+\rho^2)$, by
\begin{equation}
\Ucl(\bm{x},z)=\Po\exp\left(
i\int_{-\infty}^z\! dz'\cA^\cl_z(\bm{x},z')\right)
=\exp\Bigl(iH(r,z)\,\wh{\bm{x}}\cdot\bm{\tau}\Bigr) \ ,
\label{eq:Ucl}
\end{equation}
with
\begin{equation}
H(r,z)=\frac{r}{\sqrt{r^2+\rho^2}}\left(
\arctan\frac{z}{\sqrt{r^2+\rho^2}}+\frac{\pi}{2}\right) .
\label{eq:H(r,z)}
\end{equation}
We omit other collective coordinates than $\A(t)$ here for
simplicity.

For $U$ of \eqref{eq:U=AUclA^-1}, we have
\begin{align}
L_0&=\A\!\left[\Ucl\bigl(-i\A^{-1}\dot{\A}\bigr)\Ucl^{-1}
+i\A^{-1}\dot{\A}\right]\A^{-1} \ ,
\nn\\
L_M&=\A L^\cl_M \A^{-1} \ ,
\end{align}
and accordingly,
\begin{equation}
\tr L^5=5\,\tr\!\left(-i\A^{-1}\dot{\A} dt
\left[\left(R^\cl_M dx^M\right)^4-\left(L^\cl_M dx^M\right)^4
\right]
\right) .
\end{equation}
Here, $L^\cl$ and $R^\cl$ are given by \eqref{eq:L=-iUdU^-1} with
$U$ replaced by $\Ucl$ and $\Ucl^{-1}$, respectively.
We can show generically that, for $\Ucl$ of spherically symmetric form
\eqref{eq:Ucl} with an arbitrary $H(r,z)$ not restricted to
\eqref{eq:H(r,z)},
\begin{equation}
\left(R^\cl_M dx^M\right)^4=\left(L^\cl_M dx^M\right)^4=0 \ ,
\label{eq:(R^cldx)^4=0}
\end{equation}
and hence the WZW term of \eqref{eq:SWZW_SS} vanishes totally.

\section{New CS term}
\label{sec:NewCS}

As we saw in the last section, the CS term \eqref{eq:SCS_SS} cannot
reproduce the constraint \eqref{eq:J_8-constraint} necessary for
selecting baryon states with correct spins.
Another and potential problem about the CS term \eqref{eq:SCS_SS} is
that it is not strictly a gauge invariant quantity. Indeed, it is not
invariant under ``large'' gauge transformations (see
\eqref{eq:omega_5(A^V)}). Therefore, the physics can depend on the
choice of gauge.

\begin{figure}[htbp]
  \centering
  \epsfxsize=5cm
  \epsfbox{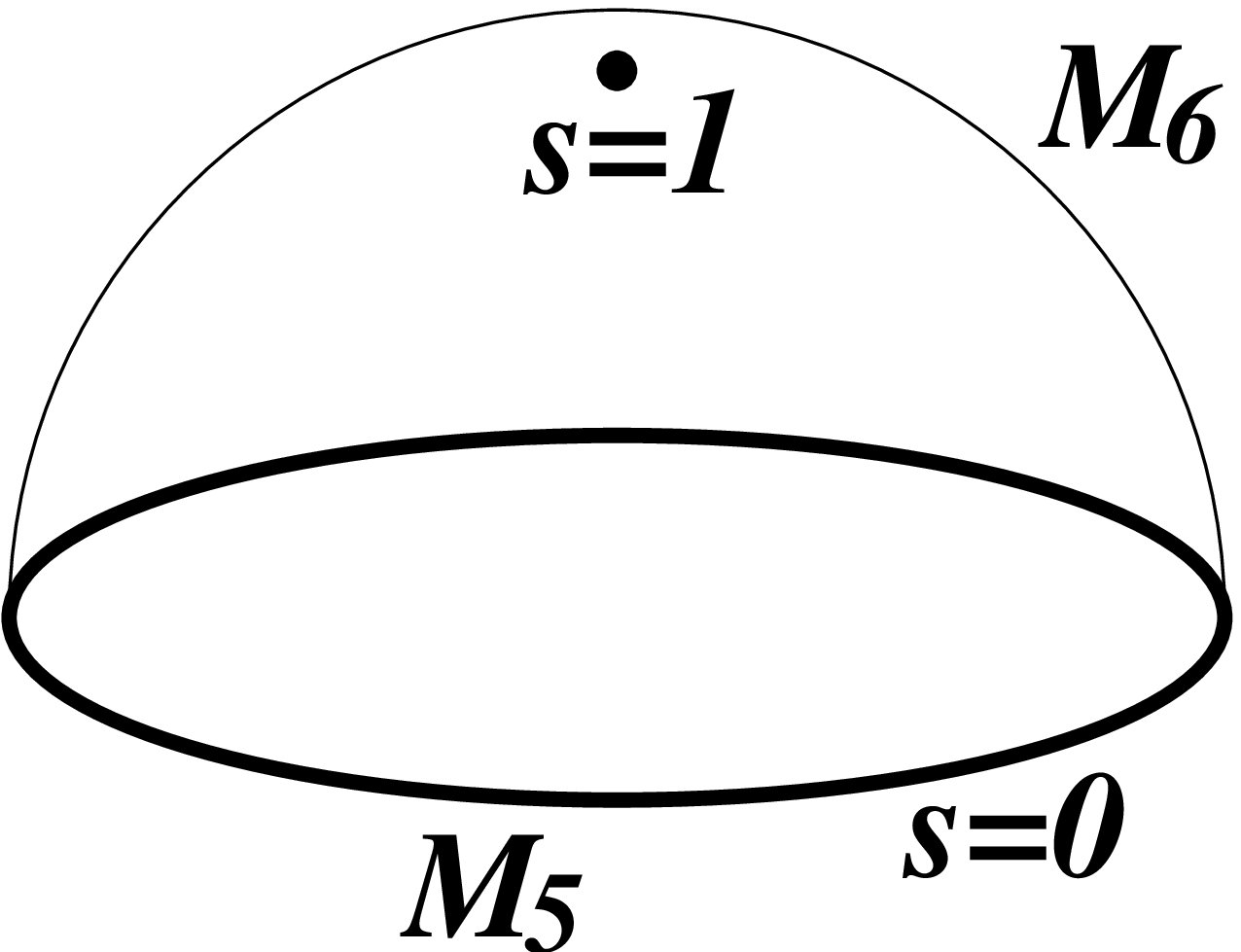}
  \caption{The space $\Msix$. The boundary $s=0$ is the original five
    dimensional spacetime $\Mfive$.}
  \label{fig:M6}
\end{figure}

To overcome these problem, we here propose another CS term for the
holographic QCD \eqref{model}. The construction is quite parallel with
that of the WZW term in the Skyrme model
\cite{Wess-Zumino,Witten:WZW}.
We introduce a new and fictitious sixth coordinate $s$ which takes
values in the interval $[0,1]$, and consider a six dimensional
spacetime $\Msix$  with coordinates
$\left(t,x^M,s\right)=\left(t,\bm{x},z,s\right)$
(see fig.\ \ref{fig:M6}).
The subspace of $s=0$ is the boundary of $\Msix$ and it is
the original five dimensional spacetime $\Mfive=\RMfour$ where the YM
action $\SYM$ \eqref{eq:SYM} is defined.
Accordingly, the gauge field on $\Msix$ has the $s$-component and is now
a function of the coordinates
$\left(t,x,s\right)=\left(t,\bm{x},z,s\right)$,
\begin{equation}
\cA(t,x,s)=\cA_0(t,x,s)dt+\cA_M(t,x,s)dx^M+\cA_s(t,x,s)ds \ ,
\label{eq:cAonMsix}
\end{equation}
and it is required to satisfy the following condition:
\begin{equation}
\cA(t,x,s=0)=\cA(t,x),\quad
\mbox{(except the $s$-component $\cA_s$)}\ .
\label{eq:condAonMsix}
\end{equation}
\begin{figure}[htbp]
  \centering
  \epsfxsize=4.5cm
  \epsfbox{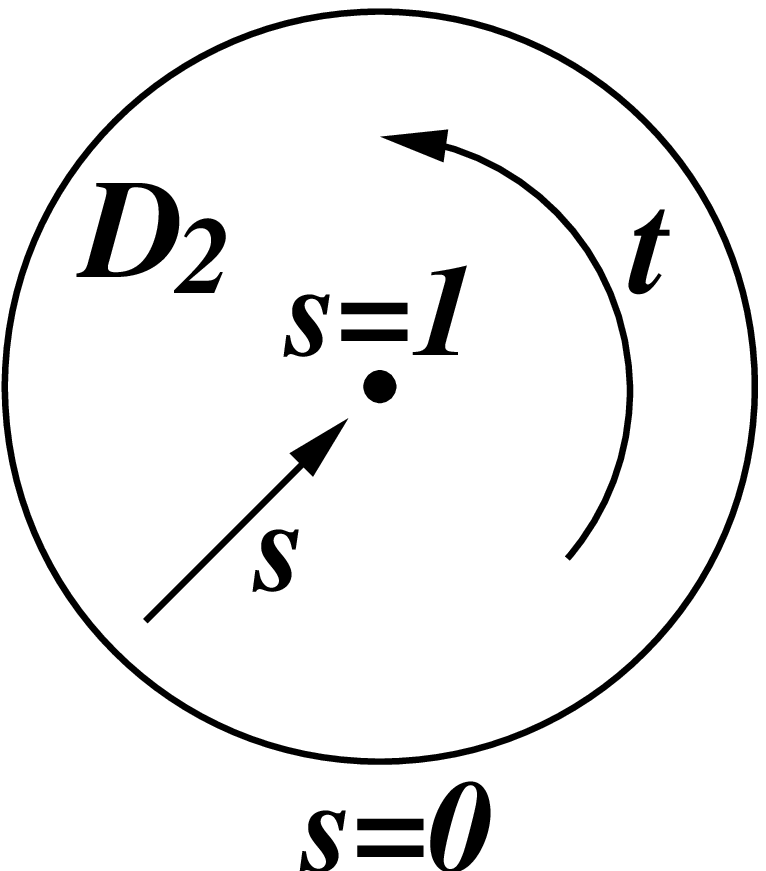}
  \caption{The space $\Dtwo$ for $(t,s)$ is a two dimensional
    disc with angle coordinate $t$ and radial one $1-s$
    ($s=0$ and $s=1$ are the boundary and the center of the
    disc, respectively).}
  \label{fig:D2}
\end{figure}
Following the case of the WZW term \cite{Jain}, we take as the space
$\Msix$ in the baryon sector the direct product
$\Msix=\Dtwo\times\Mfour$; $\Dtwo$ is the two dimensional disc for
$(t,s)$ and $\Mfour$ is for $(\bm{x},z)$.
On $\Dtwo$, $t$ is the angle coordinate and $s$ the radial one, with
$s=0$ and $s=1$ corresponding to the boundary and the center,
respectively (see fig.\ \ref{fig:D2}). Namely, we regard the space of
$t$ as $S^1$ by identifying $t=+\infty$ and $t=-\infty$.
In this case, the gauge field on $\Msix$ must respect the fact that
$s=1$ is a point on $\Dtwo$ and satisfy conditions including
\begin{equation}
\cA(t,x,s=1)=\mbox{$t$-indep.}\ .
\label{eq:cA(t,x,s=1)=t-indep.}
\end{equation}

With the above extension to the six dimensional spacetime $\Msix$, our
new CS term is given by
\begin{equation}
\newSCS=\frac{N_c}{24\pi^2}
\int_{\Msix}\tr \cF^3,
\label{eq:newSCS}
\end{equation}
where $\cF(\cA)=d\cA+i\cA^2$ is the field strength on $\Msix$
having also the $s$-component.
The ambiguity in the six dimensional extension \eqref{eq:newSCS}
is an integer times $2\pi$ and hence does not affect $\exp i\newSCS$,
as in the case of the WZW term.

Since we have
\begin{equation}
\tr\cF^3=d\omega_5(\cA) \ ,
\label{eq:trF^3=domega_5}
\end{equation}
and $\p\Msix=\Mfive$, our new
CS term \eqref{eq:newSCS} may seem merely an equivalent rewriting of
the original one \eqref{eq:SCS_SS}. This is indeed the case in the
topologically trivial sector without baryons. In the baryon sector,
however, due to the fact that we need two patches for expressing the
BPST instanton on $\Mfour(\simeq S^4)$, $\Mfive$ is not the only
boundary of $\Msix$ for gauge non-invariant quantities such as
$\omega_5$. For this reason, our new CS term can differ from the
original one in the sector with baryons. (The baryon configuration on
$\Msix$ given in this section is for the patch containing the origin
$\xi=0$. See appendix \ref{app:another} for the construction of baryon
configurations in both the patches.)

For the collective coordinate quantization of baryon using our new CS
term, we extend the gauge field \eqref{eq:cptU(3)} defined on $\Mfive$
to $\Msix$ as
\begin{equation}
\cA(t,x,s) =\bigl(\cA^\cl(x,s;X^\alpha(t,s))
+\Phi(t,x,s)dt +\Psi(t,x,s)ds\bigr)^{\A(t,s)}\ .
\label{eq:cptU(3)onMsix}
\end{equation}
Compared with \eqref{eq:cptU(3)}, the various quantities,
including $\cA^\cl$ and the collective coordinates $(\A,X^\alpha)$,
are extended to depend also on $s$, and that a new term, $\Psi ds$,
has been added.
These extensions should be done so as to satisfy the conditions
\eqref{eq:condAonMsix} and \eqref{eq:cA(t,x,s=1)=t-indep.}.
The details of the extensions are described in appendix
\ref{app:another}, and we here explain only a part necessary for the
arguments in this section.
First, the $s$-dependence of $\cA^\cl(x,s)$ should be introduced only
in the time component $\cA^\cl_0$ in such a way that the following
conditions are satisfied:
\begin{equation}
\cA^\cl_0(x,s=0)=\cA^\cl_0(x) \ ,\qquad
\cA^\cl_0(x,s=1)=0\ ,\qquad
\left[\cA^\cl_0(x,s),g(x)\right]=0 \ .
\label{eq:condcA_0(x,s)}
\end{equation}
The $s$-dependence of $\cA^\cl_0(x,s)$ can be quite arbitrary so long
as these conditions are satisfied (there is no EOM for $s\ne 0$), and
the other components $\cA^\cl_M(x)$ on $\Msix$ should not have the
explicit $s$-dependence and be the same as on $\Mfive$.
The second condition of \eqref{eq:condcA_0(x,s)} ensures that
$\newSCS[\cA^\cl]$ reproduces the same Coulomb self-energy of the
baryon solution as that from the original CS term $\SCS[\cA^\cl]$
($\newSCS[\cA^\cl]$ is reduced to the difference of $\SCS[\cA^\cl]$ at
$s=0$ and $s=1$, and the latter vanishes due to the second condition
of \eqref{eq:condcA_0(x,s)}).
The third condition of \eqref{eq:condcA_0(x,s)}, stating that
$\cA^\cl_0(x,s)$ be spanned by $\bm{1}_3$ and $t_8$, will become
necessary when we discuss the two patches in appendix
\ref{app:another}.
Other $s$-dependent quantities appearing in \eqref{eq:cptU(3)onMsix}
should of course coincide with the original ones on $\Mfive$ at $s=0$:
\begin{equation}
\A(t,s=0)=\A(t)\ ,\quad X^\alpha(t,s=0)=X^\alpha(t)\ ,
\quad \Phi(t,x,s=0)=\Phi(t,x) \ .
\label{eq:condWXPhi_s=0}
\end{equation}
We have to introduce the $\Psi ds$ term in \eqref{eq:cptU(3)onMsix}
in order to make the $s$-component of the gauge field in the patch
containing the infinity $\xi=\infty$ be regular and vanish there (see
the end of appendix \ref{subapp:M_6}).

Let us calculate our new CS term \eqref{eq:newSCS} for the baryon
configuration with collective coordinates given by
\eqref{eq:cptU(3)onMsix}. We will find that the result is just what is
necessary for reproducing the constraint \eqref{eq:J_8-constraint}.
For this purpose, we first consider $\cF$ for $\cA+\delta\cA$ with
$\delta\cA=\Phi dt + \Psi ds$, and expand it in powers of $\delta\cA$:
\begin{equation}
\cF(\cA+\delta\cA)=\cF(\cA)+D\delta\cA +i(\delta\cA)^2 \ ,
\end{equation}
with
$D\delta\cA=d\delta\cA+i(\cA\,\delta\cA+\delta\cA\,\cA)$.
This leads to
\begin{equation}
\tr\cF(\cA+\delta\cA)^3=\tr\cF(\cA)^3
+3\,d\tr\Bigl(
\delta\cA\,\cF(\cA)^2+\delta\cA\,(D\delta\cA)\,\cF(\cA)
\Bigr) \ ,
\label{eq:trF(A+dA)^3=}
\end{equation}
where we have used that
$(\delta\cA)^3=(D\delta\cA)\,(\delta\cA)^2=D\cF=0$ and
$D^2\delta\cA=i\left[\cF,\delta\cA\right]$.
Using the gauge invariance of $\newSCS$ and the formula
\eqref{eq:trF(A+dA)^3=}, $\newSCS$ for \eqref{eq:cptU(3)onMsix} is
evaluated as follows:
\begin{align}
\newSCS\bigl[&\cA=(\cA^\cl+\Phi dt+\Psi ds)^\A\bigr]
-\newSCS[\cA^\cl]
=\frac{N_c}{24\pi^2}\int_{\Msix}\!3\,d\tr\Bigl(
\delta\cA\,(\cF^\cl)^2+\delta\cA\,(D^\cl\delta\cA)\,\cF^\cl
\Bigr)
\nn\\
&=\frac{N_c}{8\pi^2}\int_{\Mfive}\tr\bigl(\Phi dt\,(F^\cl)^2\bigr)
=\frac{N_c}{2\sqrt{3}}\int\! dt\,\chi^8(t) ,
\label{eq:evalnewSCS}
\end{align}
where we have used that, on $\Mfive$ with $s=0$, $\delta\cA=\Phi dt$,
$\delta\cA\,(D^\cl\delta\cA)=0$ and $\Fh^\cl_{MN}=0$.
The last equality is nothing but \eqref{eq:intPhiF^2}.
Eq.\ \eqref{eq:evalnewSCS} implies that $\LCS$ \eqref{eq:defLCS}
for our new CS term \eqref{eq:newSCS} is
\begin{equation}
\LCS=\frac{N_c}{2\sqrt{3}}\,\chi^8(t)
=\frac{N_c}{\sqrt{3}}\tr\bigl(-i\A(t)^{-1}\dot{\A}(t)\,t_8\bigr) \ .
\label{eq:newLCS}
\end{equation}
This $\LCS$ is the same as that appears in the collective coordinate
quantization of the $SU(3)$ Skyrme model
\cite{Guadagnini,Mazur,Chemtob,Jain,Manohar}, and leads to the desired
condition \eqref{eq:J_8-constraint} (see the next section).

We should add a comment on the derivation of \eqref{eq:evalnewSCS}.
We mentioned above that $\Mfive$ is not the only boundary of
$\Msix$ for gauge non-invariant quantities. Fortunately,
$\tr\bigl(
\delta\cA\,(\cF^\cl)^2+\delta\cA\,(D^\cl\delta\cA)\,\cF^\cl\bigr)$
is gauge invariant since every constituent,
$\delta\cA$, $D^\cl\delta\cA$ and $\cF^\cl$,
transforms covariantly under the gauge transformation.
Therefore, we do not need to consider two patches for describing the
instanton. On the other hand, if we repeat the calculation of
\eqref{eq:evalnewSCS} by first using the formula
\eqref{eq:trF^3=domega_5}, we indeed need two patches since $\omega_5$
is not gauge invariant, and obtain the same result as
\eqref{eq:evalnewSCS}. Details of the calculation are given in
appendix \ref{app:another}.

\section{Quantization of the collective coordinates}
\label{sec:CCQ}

In secs.\ \ref{subsec:IntroCC} and \ref{subsec:LCC}, we introduced the
collective coordinates into the baryon solution and obtained their
lagrangian \eqref{eq:L=sumL} except the last term $\LCS$
\eqref{eq:defLCS} from the CS term.
In this section, by adopting the new CS term \eqref{eq:newSCS} and
hence $\LCS$ given by \eqref{eq:newLCS}, we will complete the
collective coordinate quantization to obtain the baryon spectra in the
three-flavor model of holographic QCD.

\subsection{Hamiltonian}

Let us start with the lagrangian of the collective coordinates
\eqref{eq:L=sumL} with $\LCS$ given by \eqref{eq:newLCS}.
This lagrangian differs from the standard collective coordinate
lagrangian of $SU(3)$ Skyrme model in that there are $L_Z$ and
$L_\rho$ terms and in that the moments of inertia, $\Iz(\rho)$ and
$\Izp(\rho)$, depends on the dynamical variable $\rho$.
However, the quantization is straightforward and we obtain the
following hamiltonian of the system (we drop the center-of-mass
coordinate $\bm{X}(t)$):
\begin{equation}
H=M_0 + H_Z + H_\rho + H_{\rho\A} \ ,
\label{eq:H=sumH}
\end{equation}
with
\begin{align}
H_Z&=-\frac{1}{2m_Z}\p_Z^2 +\frac12 m_Z\omega_Z^2 Z^2 \ ,
\label{eq:H_Z}
\\
H_\rho&=-\frac{1}{2 m_\rho}
\frac{1}{\rho^\dd}\p_\rho\bigl(\rho^\dd \p_\rho\bigr)
+\frac12 m_\rho\omega_\rho^2\rho^2 + \frac{\Q}{m_\rho \rho^2} \ ,
\label{eq:H_rho}
\\
H_{\rho\A}&=\frac{1}{2\Iz(\rho)}\sum_{a=1}^3 (J_a)^2
+\frac{1}{2\Izp(\rho)}\sum_{a=4}^7 (J_a)^2 \ .
\label{eq:H_rhoA}
\end{align}
This system must be supplemented with the constraint
\eqref{eq:J_8-constraint} coming from the fact that $\chi^8$ appears
only in $\LCS$ \eqref{eq:newLCS} in the lagrangian \eqref{eq:L=sumL}.
Here, we have taken the representation of diagonalizing $Z$ and
$\rho$.
In \eqref{eq:H_rhoA}, $J_a$ is the charge of the right $SU(3)_J$
transformation on $\A$:
\begin{equation}
[J_a,\A]=\A t_a,\qquad
[J_a,J_b]=if_{abc}J_c \ .
\end{equation}
The present system has an invariance only under the $SU(2)$ subgroup
of $SU(3)_J$, which is the group of rotation in the $\bm{x}$-space
spanned by $(J_1,J_2,J_3)$.
Besides this, our system has the full invariance under the $SU(3)_I$
flavor transformation. The charge $I_a$ of $SU(3)_I$ satisfies
\begin{equation}
[I_a,\A]=-t_a\A,\qquad
[I_a,I_b]=if_{abc}I_c,\qquad
[I_a,J_b]=0 \ .
\end{equation}
Since the relation $I=\A J\A^{-1}$ holds for $I=I_a t_a$ and
$J=J_at_a$, we have $\tr I^2=\tr J^2$ and $\tr I^3=\tr J^3$.
Therefore, the representation of $SU(3)_I$ and $SU(3)_J$ must be the
same.

The first term of \eqref{eq:H_rho} is chosen so that it is
hermitian with respect to the inner-product
$(f,g)=\int_0^\infty d\rho \rho^\dd f^*(\rho) g(\rho)$.
In the $N_f=2$ case of \cite{HSSY}, we had $\eta=3$ since
we identified $\rho$ and $\A$ as the radial coordinate and the
orientation, respectively, of the part of the instanton moduli space
$\bR^4/\bZ_2$ with line element
$(\delta s)^2=\rho^2\frac12\tr(-i\A^{-1}\delta\A)^2+(\delta\rho)^2$.
In the present $N_f=3$ case, it is natural to put $\dd=8$.
However, we leave $\dd$ generic until we compare our result on the
baryon spectra with experimental data.

\subsection{Baryon mass formula}

Let us solve the Schr\"odinger equation of our collective
coordinate system to obtain the spectra.
First, we consider the hamiltonian $H_\rho+H_{\rho\A}$ by
taking the $(p,q)$ representation for the two $SU(3)$, $SU(3)_J$ and
$SU(3)_I$.
For a state in this representation and with spin $j$,  we have
\begin{align}
\sum_{a=1}^8(J_a)^2&=\frac13\left(p^2+q^2+pq+3(p+q)\right) ,
\\
\sum_{a=1}^3(J_a)^2&=j(j+1) \ ,
\end{align}
and the $\rho$ part of the hamiltonian $H_\rho + H_{\rho\A}$ becomes
\begin{equation}
\Hrhotot=-\frac{1}{2 m_\rho}
\frac{1}{\rho^\dd}\p_\rho\bigl(\rho^\dd\p_\rho\bigr)
+\frac12 m_\rho\omega_\rho^2\rho^2 + \frac{\Q'}{m_\rho\rho^2} \ ,
\end{equation}
where $\Q'$ is the sum of $\Q$ and the contribution from $H_{\rho\A}$:
\begin{equation}
\Q'=\frac{N_c^2}{15}
+\frac43\left(p^2+q^2+pq+3(p+q)\right)-2j(j+1) \ .
\end{equation}
The first term $N_c^2/15$ is the sum of $K=(2/5)N_c^2$ and
$-N_c^2/3$ coming from $-(J_8)^2/(2\Izp(\rho))$ with $J_8$ given by
\eqref{eq:J_8-constraint}.

Now we consider solving the Schr\"odinger equation
\begin{equation}
\Hrhotot\psi(\rho)=E_{\rho\A} \psi(\rho) \ .
\end{equation}
This equation is reduced via
\begin{equation}
\psi(\rho)=e^{-z/2} z^\beta v(z) \ ,
\end{equation}
with
\begin{equation}
z=m_\rho\omega_\rho\rho^2,\qquad
\beta=\frac14\left(\sqrt{(\dd-1)^2+8\Q'}-(\dd-1)\right) ,
\end{equation}
to a confluent hypergeometric differential equation for $v(z)$:
\begin{equation}
\left\{z\drv{^2}{z^2}
+\left(2\beta+\frac{\dd+1}{2}-z\right)\drv{}{z}
+\left(\frac{E_{\rho\A}}{2\omega_\rho}-\beta-\frac{\dd+1}{4}
\right)\right\}v(z) = 0 \ .
\label{eq:DEpsi}
\end{equation}
A normalizable regular solution to \eqref{eq:DEpsi} exists only when
$E_{\rho\A}/(2\omega_\rho)-\beta-(\dd+1)/4=n_\rho=0,1,2,3,\cdots$.
Namely, the energy eigenvalues are given by
\begin{equation}
E_{\rho\A}=\omega_\rho\left(2 n_\rho
+\frac12\sqrt{(\dd-1)^2+8\Q'}+1\right) .
\label{eq:E_rhoA=}
\end{equation}
The eigenvalues of the $Z$ part hamiltonian $H_Z$ \eqref{eq:H_Z} are
simply those of a harmonic oscillator:
\begin{equation}
E_Z=\omega_Z\left(n_Z+\frac12\right),\quad
(n_Z=0,1,2,3,\cdots) \ ,
\label{eq:E_Z=}
\end{equation}
Adding \eqref{eq:E_rhoA=} and \eqref{eq:E_Z=}, the baryon mass
formula in the present model is finally given by
\begin{equation}
M=M_0+\sqrt{\frac{(\dd-1)^2}{24}+\frac{\Q'}{3}}\,
+\sqrt{\frac23}\left(n_\rho+n_Z+1\right) .
\label{eq:BS}
\end{equation}

In the above arguments, $N_c$ was arbitrary and we had not imposed the
constraint \eqref{eq:J_8-constraint} on the states specified by $(p,q)$
and $j$.
Putting $N_c=3$, the constraint \eqref{eq:J_8-constraint},
\begin{equation}
J_8=\frac{\sqrt{3}}{2} \ ,
\label{eq:J_8=sqrt3/2}
\end{equation}
implies that $(p,q)$ must satisfy
\begin{equation}
p+ 2 q= 3\times(\mbox{integer}) \ .
\label{eq:p+2q=3n}
\end{equation}
The allowed states with smaller $(p,q)$ satisfying the constraints
\eqref{eq:p+2q=3n} and \eqref{eq:J_8=sqrt3/2} as well as their $\Q'$
values are as follows:
\begin{align}
(p,q) &= (1,1),\quad j=\frac12,\quad \Q'=\frac{111}{10},
\quad \text{(octet)}
\nn\\
(p,q) &= (3,0),\quad j=\frac32, \quad \Q'=\frac{171}{10},
\quad \text{(decuplet)}
\nn\\
(p,q) &= (0,3),\quad j=\frac12, \quad \Q'=\frac{231}{10},
\quad \text{(anti-decuplet)} \ .
\label{eq:valuesQ'}
\end{align}

\subsection{Comparison with experimental data}

The present three-flavor holographic QCD model is not a realistic one
since all the quarks are massless.
It does not make much sense to compare seriously the obtained baryon
spectrum \eqref{eq:BS} with experimental data unless we add at least
to the strange quark a mass to break the $SU(3)_I$ symmetry
(see refs.\ \cite{HHM,Evans,BSS,Dhar} for attempts to introduce quark
masses in the SS-model).
Below we will make comparison of our baryon mass formula \eqref{eq:BS}
with the observed spectra of baryons.
However, we keep our analysis very short for this reason.

{}From \eqref{eq:BS} with $\eta=8$, the mass difference between the
octet and the decuplet baryons with the same $(n_\rho,n_Z)$, and that
between the octet and the anti-decuplet are given in units of
$\MKK$ as follows:
\begin{align}
M_{\bm{10}}-M_{\bm{8}}&=0.386208 \ ,
\\
M_{\bm{10}^*}-M_{\bm{8}}&=0.724987 \ .
\label{eq:M10*-M8}
\end{align}
The value of $M_{\bm{10}}-M_{\bm{8}}$ is much smaller (nearly $64\%$)
than the corresponding value ($M_{l=3}-M_{l=1}=0.600$) in the $N_f=2$
case \cite{HSSY}. Therefore, the favored value of $\MKK$ for
realizing the experimental data
$M^{\rm exp}_{\bm{10}}-M^{\rm exp}_{\bm{8}}=(1232-940)\MeV=292\,\MeV$
of low-lying non-strange baryons is
\begin{equation}
\MKK=756\,\MeV \ .
\label{eq:MKK}
\end{equation}
This is smaller than $\MKK=949\,\MeV$
determined from the $\rho$ meson mass \cite{SaSu1,SaSu2}, but is
large than $\MKK\simeq 500\,\MeV$ in the $N_f=2$ case
\cite{HSSY}.
The dependence of the mass formula \eqref{eq:BS} on $(n_\rho,n_Z)$ is
the same as in the $N_f=2$ case (see eq.\ (5.31) of ref. \cite{HSSY}).
Therefore, \eqref{eq:BS} with $\MKK$ given by \eqref{eq:MKK}
predicts heavier masses for the excited baryon states than in
\cite{HSSY}, though the comparison with experimental data is not so
bad.
Finally, adopting the value \eqref{eq:MKK} for $\MKK$,
eq.\ \eqref{eq:M10*-M8} for the anti-decuplet predicts
\begin{equation}
M_{\bm{10}^*}-M_{\bm{8}}=548\,\MeV \ .
\end{equation}
This is close to the experimental value
$M^{\rm exp}_{\bm{10}^*}-M^{\rm exp}_{\bm{8}}
=(1530-940)\MeV=590\,\MeV$ obtained using the reported $\Theta^+$ mass
of $1530\,\MeV$ \cite{PDG}. Of course, we cannot take this result
seriously due to the lack of strange quark mass in our model.

\section{Summary and discussions}
\label{sec:summary}

In this paper, we studied baryons in the SS-model with three flavors.
The baryon solution is given by an $SU(3)$ embedding of the BPST
instanton solution with small size of order $\lambda^{-1/2}$,
and we carried out the collective coordinate quantization of the
baryon solution.
Although our analysis is quite parallel with the previous one
for the two flavor case \cite{HSSY}, the three flavor case is the
first nontrivial place where the non-abelian part of the CS term
should play a critical role of giving the constraint which selects
baryons with correct spins.
We found that the original CS term \eqref{eq:SCS_SS} given in
terms of the CS 5-form does not work, and proposed another CS term
\eqref{eq:newSCS} by introducing the fictitious sixth coordinate $s$.
These two CS terms are naively equivalent, but they are different ones
in the baryon sector which cannot be described only by one patch.
In fact, we found that our new CS term leads to the desired
constraint. Using our new CS term, we completed the collective
coordinate quantization and obtained the baryon mass formula
\eqref{eq:BS}.
The $N$-$\Delta$ mass difference favors the value of $\MKK$ which is
larger than that in the $SU(2)$ case \cite{HSSY} but is smaller than
that determined by the $\rho$ meson mass \cite{SaSu1,SaSu2}.
Of course, serious comparison of our mass formula with experimental
data is meaningless since all the quarks are massless in the present
model.

We finish this paper by discussing remaining problems in the three
flavor SS-model, especially concerning the CS term.
First is the origin of the sixth coordinate $s$ for expressing our CS
term \eqref{eq:newSCS}. In this paper, the coordinate $s$ was
introduced simply by hand just like the fifth coordinate in the WZW
term. However, recall that the original CS term \eqref{eq:SCS_SS} has
been obtained from the following coupling:
\begin{equation}
\SCS^{\rm D8}=\frac{1}{48\pi^3}\int_{D8}C_3\tr \cF^3 \ ,
\label{eq:S^D8}
\end{equation}
where the integration is over the D8-brane, and $C_3$ is RR 3-form of
the D4-brane background. Eq.\ \eqref{eq:S^D8} vanishes identically if
we consider only $A_0$ and $A_M$ ($M=1,2,3,z$) on $D8$ depending
only on $(t,x^M)$.  Therefore, in ref.\ \cite{SaSu1}, they adopted
\eqref{eq:SCS_SS} which is obtained from \eqref{eq:S^D8} by carrying
out the integration-by-parts using the formula
\eqref{eq:trF^3=domega_5}, discarding the surface term, and then using
$1/(2\pi)\int_{S^4}d C_3=N_c$.
It would be interesting if we could directly relate our CS term
\eqref{eq:newSCS} with \eqref{eq:S^D8} and find a ``physical
origin'' of the sixth coordinate $s$.
We cannot, however, adopt \eqref{eq:S^D8} itself instead of our
\eqref{eq:newSCS} for a number of reasons. For example, if we allow a
gauge field component other than $A_0$ and $A_M$ for \eqref{eq:S^D8},
it must be contained also in the YM action $\SYM$.

The second problem is on the reproducibility of chiral anomaly in
QCD in the presence of the background gauge field defined by
$A_{L/R}(t,\bm{x})=\lim_{z\to +\infty/-\infty}A(t,\bm{x},z)$.
The chiral anomaly is correctly reproduced from the original CS term
\eqref{eq:SCS_SS} using the gauge transformation property
\eqref{eq:omega_5(A^V)} of $\omega_5(\cA)$ \cite{SaSu1}.
On the other hand, if we adopt our new CS term $\newSCS$
\eqref{eq:newSCS}, anomaly seems not to arise at all since
\eqref{eq:newSCS} is strictly gauge invariant. A quick remedy to this
problem is to add to $\newSCS$ the following boundary term
\begin{equation}
\Delta S_{\rm CS}=-\frac{N_c}{24\pi^2}\left(
\int_{Z_+} - \int_{Z_-}\right)\omega_5(\cA) \ ,
\label{eq:DeltaSCS}
\end{equation}
where the integration region $Z_\pm$ is the $z=\pm\infty$ boundary of
$\Msix$. Note that $\Delta S_{\rm CS}$ vanishes in the absence of the
background gauge fields $A_{L/R}$ since we have $Z_+=Z_-$ in this
case. The modified CS term $\newSCS+\Delta S_{\rm CS}$ reproduces the
chiral anomaly at least in the sector without baryons.
It would be desirable to find a more concise definition of the CS term
which can reproduce both the constraint \eqref{eq:J_8-constraint} and
the chiral anomaly.

Finally, for serious comparison of our result, in particular, the
baryon mass formula \eqref{eq:BS}, with experiments, we have to redo
the analysis by introducing the strange quark mass. This is the most
important subject for the three flavor SS-model.

\section*{Acknowledgements}
We would like to thank T.~Eguchi, A.~Miwa, T.~Sakai, S.~Seki,
S.~Sugimoto, S.~Terashima and S.~Yamato for valuable discussions.
We also thank the Yukawa Institute for Theoretical
Physics at Kyoto University. Discussions during the YITP workshop
YITP-W-07-05 on ``String Theory and Quantum Field Theory''
were useful to complete this work.
The work of H.~H.\ was supported in part by a Grant-in-Aid for
Scientific Research (C) No.\ 18540266 from the Japan Society for the
Promotion of Science (JSPS).

\appendix

\section{Determination of $\bm{\Phi(t,x)}$}
\label{derive:Phi}

In this appendix, we solve \eqref{Gausslaw2} to obtain $\Phi$
\eqref{sol:Phi} in the $SU(3)$ case. We essentially follow appendix A
of \cite{HSSY}. Let us decompose $\Phi$ into three parts, each of
which depends on the time derivative of one of the three kinds of
collective coordinates:
\begin{equation}
\Phi = \Phi_X + \Phi_\rho + \Phi_{SU(3)} \ .
\end{equation}
Then, \eqref{Gausslaw2} is reduced to the following three equations:
\begin{align}
&D_M^{\rm cl}\left(\dot{X}^N\frac{\p}{\p X^N}A^{\rm cl}_M
- D_M^{\rm cl}\Phi_X\right) = 0 \ ,
\label{eq:EqPhiX}\\
&D_M^{\rm cl}\left(\dot\rho\frac{\p}{\p\rho}A_M^{\rm cl}
- D_M^{\rm cl}\Phi_\rho\right) = 0 \ ,
\label{eq:EqPhirho}\\
&D_M^{\rm cl}D_M^{\rm cl}\Phi_{SU(3)} = 0 \ .
\label{eq:EqPhiSU(3)}
\end{align}
Solutions to eqs.\ \eqref{eq:EqPhiX} and \eqref{eq:EqPhirho} are the
same as in the $SU(2)$ case of \cite{HSSY}:
\begin{equation}
\Phi_X = -\dot{X}^MA_M^{\rm cl} \ ,
\qquad
\Phi_\rho = 0 \ .
\end{equation}
To solve \eqref{eq:EqPhiSU(3)}, it is convenient work in the singular
gauge, namely, the gauge where the BPST solution is singular at the
origin but is regular at the infinity.
Let us specify the quantities in the singular gauge by attaching the
overline on the corresponding one in the regular gauge.
The BPST solution in the singular gauge is related to \eqref{eq:BPST}
in the regular gauge via the gauge transformation by $g(x)^{-1}$,
\begin{equation}
\ol{A}^\cl_M(x)=g(x)^{-1}\left(A^\cl(x)-i\p_M\right)g(x)
= -i\left(1-f(\xi)\right)g(x)^{-1}\p_M g(x) \ .
\label{eq:BPSTbar}
\end{equation}
Since $\Phi_{SU(3)}$ transforms covariantly under the gauge
transformation, we have
\begin{equation}
\ol{\Phi}_{SU(3)}(t,x)=g(x;X(t))^{-1}\Phi_{SU(3)}(t,x)g(x;X(t)) \ ,
\end{equation}
and eq.\ \eqref{eq:EqPhiSU(3)} in the singular gauge is
\begin{equation}
\ol{D}^\cl_M\ol{D}^\cl_M\ol{\Phi}_{SU(3)}=0 \ .
\end{equation}
This equation is reduced, by assuming the form
\begin{equation}
\ol{\Phi}_{SU(3)} = u^a(\xi)t_a \ ,
\label{eq:olPhi=u^at_a}
\end{equation}
and using the properties $\p_M\ol{A}^\cl_M=0$ and
$(x-X)^M\ol{A}^\cl_M=0$,
to the following differential equation for each $u^a(\xi)$:
\begin{equation}
\frac1{\xi^3}\drv{}{\xi}\left(\xi^3\drv{}{\xi}u^a(\xi)\right)
=C_a\frac{\left(1-f(\xi)\right)^2}{\xi^2}\,u^a(\xi) \ ,
\label{eq:DEu^a}
\end{equation}
where $C_a$ is defined in terms of the structure constant $f_{abc}$
of $SU(3)$ by
$4\sum_{c=1}^3\sum_{d=1}^8f_{acd}f_{bcd}=\delta_{ab}C_a$, and it is
given concretely by
\begin{equation}
C_a=
\begin{cases}
8 & (a=1,2,3)\\ 3 &(a=4,5,6,7) \\ 0 &(a=8)
\end{cases}.
\end{equation}
The solution to \eqref{eq:DEu^a} regular at $\xi=0$ is
\begin{equation}
u^a(\xi) =
\begin{cases}
f(\xi) &(a=1,2,3)\\
f(\xi)^{1/2} &(a=4,5,6,7)\\
1 &(a=8)
\end{cases}
,
\label{eq:u^a}
\end{equation}
up to a multiplicative constant for each $u^a$.
Back to the regular gauge, we find that the general solution to
\eqref{eq:EqPhiSU(3)} is
\begin{equation}
\Phi_{SU(3)}(t,x)=\chi^a(t)\Phi_a(x;X^\alpha(t)) \ ,
\end{equation}
with $\Phi_a$ given by
\begin{equation}
\Phi_a(x;X^\alpha(t))=u^a(\xi)g(x;X(t))t_a g(x;X(t))^{-1} \ ,
\label{sol:Phia}
\end{equation}
and $\chi^a(t)$ being arbitrary functions of $t$ only.
Note that $X^\alpha=(X^M,\rho)$ in $u^a(\xi)$ is also made
time-dependent.

If we had solved \eqref{eq:EqPhiSU(3)} in the regular gauge by
assuming \eqref{eq:olPhi=u^at_a} for $\Phi_{SU(3)}$, we would have
obtained \eqref{eq:DEu^a} with $1-f$ replaced by $f$.
However, its solutions are divergent either at $\xi=0$ or at
$\xi=\infty$.

\section{Formulas of $\bm{\omega_5}$}
\label{app:formulas_omega_5}

Here, we summarize the formulas related with $\omega_5(\cA)$
\eqref{eq:omega_5} (the gauge group can be arbitrary).
First, under the gauge transformation
$\cA\to\cA^V=V\left(\cA-id\right)V^{-1}$, we have
\begin{equation}
\omega_5(\cA^V)=\omega_5(\cA)+\frac{1}{10}\tr L^5+d\alpha_4(L,\cA)
\ ,
\label{eq:omega_5(A^V)}
\end{equation}
with $\alpha(L,\cA)$ defined by
\begin{equation}
\alpha_4(L,\cA)=\frac12\tr\!\left[
L\left(\cA\cF+\cF\cA-i\cA^3\right)+\frac{i}{2} L\cA L\cA
-i L^3\cA\right],\quad
\left(L=-iV^{-1}dV\right)\ .
\label{eq:alpha_4}
\end{equation}
Second, the change of $\omega_5(\cA)$ under an arbitrary infinitesimal
deformation $\cA\to\cA+\delta\cA$ is
\begin{equation}
\omega_5(\cA+\delta\cA)=\omega_5(\cA)
+3\tr\bigl(\delta\cA\,\cF^2\bigr)+d\beta(\delta\cA,\cA)
+O\bigl((\delta\cA)^2\bigr)\ ,
\label{eq:omega_5(A+deltaA)}
\end{equation}
where $\beta(\delta\cA,\cA)$ is
\begin{equation}
\beta(\delta\cA,\cA)=\tr\!\left[\delta\cA\left(
\cF\cA+\cA\cF-\frac{i}{2}\cA^3\right)\right]\ .
\label{eq:beta}
\end{equation}

\section{Another derivation of (\ref{eq:evalnewSCS})}
\label{app:another}

In this appendix, we present another way of deriving the result of
\eqref{eq:evalnewSCS}: We reduce \eqref{eq:newSCS} to surface
integrations by using \eqref{eq:trF^3=domega_5}, but taking into
account that $\Mfive$ is not the unique boundary of $\Msix$ for
$\omega_5(\cA)$. For this purpose, we first define the gauge fields on
the two patches in $\Mfive$ (appendix \ref{subapp:M_5}) and in $\Msix$
(appendix \ref{subapp:M_6}). Rederivation of \eqref{eq:evalnewSCS} is
done in appendix \ref{subapp:rederiv}.

\subsection{Baryon configurations
on the two patches in $\bm{\Mfive}$}
\label{subapp:M_5}

\begin{figure}[htbp]
  \centering
  \epsfxsize=5.7cm
  \epsfbox{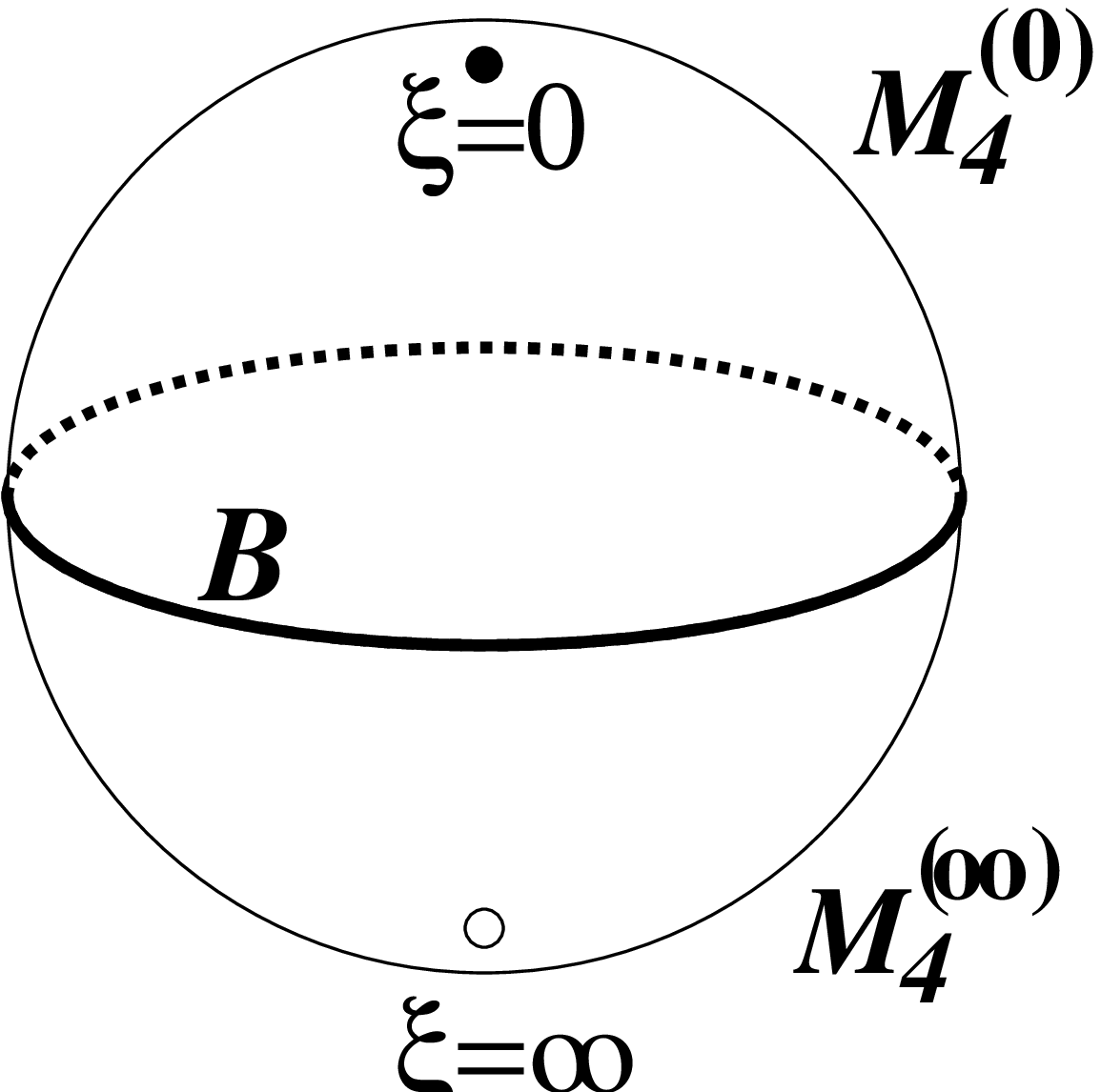}
  \caption{The space $\Mfour (\simeq S^4)$ for $x^M=(\bm{x},z)$ in the
    baryon sector consists of two patches, $\Mfourzero$ and
    $\Mfourinfty$, which are separated by the boundary $B$.}
  \label{fig:M4withB}
\end{figure}

First of all, we need two patches for describing the baryon
solution (BPST solution) in the whole of $\Mfour (\simeq S^4)$
including both the origin $\xi=0$ and the infinity $\xi=\infty$
\cite{EGH}.
Let $\Mfourzero$ and $\Mfourinfty$ be the patches containing the
origin and the infinity, respectively, separated by the boundary $B$;
$\Mfour=\Mfourzero+\Mfourinfty$ and $\p\Mfourzero=-\p\Mfourinfty=B$
(see fig.\ \ref{fig:M4withB}).
In the patch $\Mfourzero$, we adopt the BPST solution $A^\cl_M$
\eqref{eq:BPST}, while in the other patch $\Mfourinfty$ we use
$\ol{A}^\cl_M$ \eqref{eq:BPSTbar} in the singular gauge.
These two are related via the $SU(2)$ gauge transformation by
$g(x)^{-1}$.
The time-components of the solution, $\wh{A}^\cl_0$ \eqref{A0hat} and
$A^\cl_0$ \eqref{A0}, are common between the two patches since they
are $SU(2)$ invariant, and they are indeed regular both at the origin
and the infinity. Summarizing, the $U(3)$ classical solutions,
$\cA^\cl(x)$ in $\Mfourzero$ and $\ol{\cA}^\cl(x)$ in $\Mfourinfty$,
are related as a whole via the gauge transformation by $g(x)^{-1}$:
\begin{equation}
\ol{\cA}^\cl(x)=\bigl(\cA^\cl\bigr)^{g(x)^{-1}}(x)
=g(x)\bigl(\cA^\cl(x)-id\bigr)g(x)^{-1} \ .
\label{eq:olcAcl=cAcl^g^-1}
\end{equation}

The baryon configuration $\cA(t,x)$ with collective coordinates on
$\Mfivezero=\bR\times\Mfourzero$ is given by \eqref{eq:cptU(3)}.
As seen from the arguments in appendix \ref{derive:Phi},
the corresponding one, $\ol{\cA}(t,x)$, on
$\Mfiveinfty=\bR\times\Mfourinfty$ is given by\footnote{
Note that the gauge transformation $\cA^V=V(\cA-id)V^{-1}$ on $\cA$
has the property $\cA^{V_1 V_2}=\bigl(\cA^{V_2}\bigr)^{V_1}$.
}
\begin{equation}
\ol{\cA}(t,x)=\cA^{\A(t)g(x;X(t))^{-1}\A(t)^{-1}}(t,x)
=\bigl(\ol{\cA}^\cl(x;X^\alpha(t))+\ol{\Phi}(t,x)dt\bigr)^{\A(t)}\ ,
\label{eq:olcA(t,x)}
\end{equation}
with
\begin{equation}
\ol{\Phi}(t,x)=g(x;X(t))\bigl(\Phi(t,x)-i\p_0\bigr)g(x;X(t))^{-1}
=-\dot{X}^N(t)\ol{A}^\cl_N(x;X^\alpha(t))
+\sum_{a=1}^8\chi^a(t)u^a(\xi)t_a \ .
\end{equation}
Note that all the components of $\ol{\cA}(t,x)$ vanish sufficiently
fast at the infinity $\xi=\infty$. In particular, we have
$\ol{A}_0(t,x)=\cO(1/\xi^2)$ as $\xi\to\infty$ due to that
$u^a(\xi)=1+\cO(1/\xi^2)$ for all $a$.
Therefore, our $\ol{\cA}(t,x)$ is indeed well-defined on $\Mfourinfty$
containing the infinity.

\subsection{Baryon configurations on $\bm{\Msix}$}
\label{subapp:M_6}

Then, we have to extend the baryon configurations on
$\Mfive=\Mfivezero+\Mfiveinfty$ to
$\Msix=\Dtwo\times\Mfour=\Msixzero+\Msixinfty$
with $\Msix^{(0/\infty)}=\Dtwo\times\Mfour^{(0/\infty)}$
for our new CS term \eqref{eq:newSCS}.
Recall that $\Dtwo$ is the disc with angle coordinate $t$ and radial
one $s$, and $s=0$ and $s=1$ correspond to the circumference and the
center of the disc, respectively (fig.\ \ref{fig:D2}).

The baryon configuration $\cA(t,x,s)$ \eqref{eq:cAonMsix} on
$\Msixzero=\Dtwo\times\Mfourzero$ must satisfy the condition
\eqref{eq:condAonMsix} at $s=0$. In addition,  it must respect the
fact that $s=1$ is a point on $\Dtwo$ and satisfy the conditions
including \eqref{eq:cA(t,x,s=1)=t-indep.}.
Concretely, $\cA(t,x,s)$ is given by \eqref{eq:cptU(3)onMsix}
in terms of $s$-dependent collective coordinates
$(\A(t,s),X^\alpha(t,s))$ as well as $\Phi(t,x,s)$ and $\Psi(t,x,s)$
which satisfy the conditions \eqref{eq:condWXPhi_s=0} at $s=0$ and the
following ones at $s=1$ necessary for $s=1$ to be a point on $\Dtwo$:
\begin{equation}
\cO(t,x,s=1)=\mbox{$t$-indep.},\quad
\p_s\cO(t,x,s)\bigr|_{s=1}=0,\quad
\left(\cO=\A, X^\alpha, \Phi, \Psi\right) ,
\end{equation}
As we explained below \eqref{eq:cptU(3)onMsix}, the classical
configuration $\cA^\cl$ in \eqref{eq:cAonMsix} is given by
$\cA^\cl(x,s)=\cA^\cl_0(x,s)dt + \cA^\cl_M(x)dx^M$
with $s$-dependent $\cA^\cl_0(x,s)$ satisfying the condition
\eqref{eq:condcA_0(x,s)}.

The baryon configuration $\ol{\cA}(t,x,s)$ on the other patch
$\Msixinfty=\Dtwo\times\Mfourinfty$, which is as an extension of
\eqref{eq:olcA(t,x)}, is given by
\begin{align}
\ol{\cA}(t,x,s)&=\cA^{\A(t,s)g(x;X(t,s))^{-1}\A(t,s)^{-1}}(t,x,s)
\nn\\
&=\bigl(\ol{\cA}^\cl(x,s;X^\alpha(t,s))
+\ol{\Phi}(t,x,s)dt +\ol{\Psi}(t,x,s)ds\bigr)^{\A(t,s)}\ .
\label{eq:olcA(t,x,s)}
\end{align}
This extension should satisfy
\begin{equation}
\ol{\cA}(t,x,s=0)=\ol{\cA}(t,x),\quad
\ol{\cA}(t,x,s=1)=\mbox{$t$-indep.},\quad
\ol{\cA}(t,x,s)\underset{\xi\to\infty}{\longrightarrow}0 \ .
\end{equation}
The precise meaning of the third condition is that $\ol{\cA}$ tends to
zero faster than $\cO(1/\xi)$.
The classical configuration $\ol{\cA}^\cl$ in \eqref{eq:olcA(t,x,s)}
is given by
$\ol{\cA}^\cl(x,s)=\cA^\cl_0(x,s)dt +\ol{\cA}^\cl_M(x)dx^M$
in terms of the same $\cA^\cl_0(x,s)$ as in $\cA^\cl(x,s)$ on
$\Msixzero$. Owing to the third condition of \eqref{eq:condcA_0(x,s)},
eq.\ \eqref{eq:olcAcl=cAcl^g^-1} continues to hold on $\Msix$:
\begin{equation}
\ol{\cA}^\cl(x,s)=\bigl(\cA^\cl\bigr)^{g(x)^{-1}}(x,s)
=g(x)\bigl(\cA^\cl(x,s)-id\bigr)g(x)^{-1} \ .
\label{eq:olcAcl=cAcl^g^-1_Msix}
\end{equation}
Note that the following relations hold:
\begin{align}
\ol{\Phi}(t,x,s)
&=g(x;X(t,s))^{-1}\bigl(\Phi(t,x,s)-i\p_0\bigr)g(x;X(t,s)) \ ,
\nn\\
\ol{\Psi}(t,x,s)
&=g(x;X(t,s))^{-1}\bigl(\Psi(t,x,s)-i\p_s\bigr)g(x;X(t,s)) \ .
\end{align}

Our $\newSCS$ is independent of the details of extending the
various quantities into $\Msix$.
In particular, $\Phi(t,x,s)$ and $\Psi(t,x,s)$ for $s\ne 0$ are
subject to no restrictions of the Gauss law constraint, and hence they
are not uniquely determined.
An example of $\Phi(t,x,s)$ and $\Psi(t,x,s)$ is
\begin{align}
\Phi(t,x,s)&=-\dot{X}^N(t,s)A^\cl_N(x;X^\alpha(t,s))
\nn\\
&\qquad
-2i\sum_{a=1}^8u^a(\xi)
\tr\left[t_a\A(t,s)^{-1}\p_0\A(t,s)\right]
g(x;X(t,s))t_a g(x;X(t,s))^{-1} \ ,
\label{eq:Phi(t,x,s)}
\\
\Psi(t,x,s)&=-\p_s X^N(t,s)A^\cl_N(x;X^\alpha(t,s))
\nn\\
&\qquad
-2i\sum_{a=1}^8u^a(\xi)
\tr\left[t_a\A(t,s)^{-1}\p_s\A(t,s)\right]
g(x;X(t,s))t_a g(x;X(t,s))^{-1} \ .
\label{eq:Psi(t,x,s)}
\end{align}
The corresponding $\ol{\Phi}(t,x,s)$ and $\ol{\Psi}(t,x,s)$
are obtained from \eqref{eq:Phi(t,x,s)} and \eqref{eq:Psi(t,x,s)},
respectively, by replacing $A^\cl_N$ with $\ol{A}_N^\cl$ and
removing $g(x;X(t,s))$.
Note that we have $\ol{A}_s(t,x,s)=\cO(1/\xi^2)$ $(\xi\to\infty$) for
the present $\ol{\Psi}$.

\subsection{Rederivation of (\ref{eq:evalnewSCS})}
\label{subapp:rederiv}

Having finished the preparation, let us turn to the evaluation of
$\newSCS$ \eqref{eq:newSCS} by reducing it to surface integrations.
Taking $\cA(t,x,s)$ \eqref{eq:cptU(3)onMsix} and $\ol{\cA}(t,x,s)$
\eqref{eq:olcA(t,x,s)} as the gauge field on $\Msixzero$ and
$\Msixinfty$, respectively, and using that
$\p\Msixzero=\Mfivezero +\Dtwo\times B$
and $\p\Msixinfty=\Mfiveinfty-\Dtwo\times B$, we obtain
\begin{align}
\int_{\Msix}\tr\left(\cF^3 -(\cF^\cl)^3\right)&=
\int_{\Mfivezero}\left(\omega_5(\cA)-\omega_5(\cA^\cl)\right)
+\int_{\Mfiveinfty}\left(\omega_5(\ol{\cA})
-\omega_5(\ol{\cA}^\cl)\right)
\nn\\
&\qquad
+\int_{\Dtwo\times B}
\left[\left(\omega_5(\cA)-\omega_5(\ol{\cA})\right)
-\left(\omega_5(\cA^\cl)-\omega_5(\ol{\cA}^\cl)\right)\right] \ .
\label{eq:rederivI}
\end{align}
Then, recall \eqref{eq:intomega5(cA)=intomega5(cA^cl)},
stating that the original CS term \eqref{eq:SCS_SS} does not depend on
the collective coordinates at all.
In quite the same manner, calculation in the singular gauge leads to
\begin{equation}
\int_{\Mfive=\Mfivezero+\Mfiveinfty}\left(
\omega_5(\ol{\cA})-\omega_5(\ol{\cA}^\cl)\right)=0 \ .
\label{eq:int(omega-omega)=0}
\end{equation}
Using this and the formula \eqref{eq:omega_5(A^V)} with
$V=\A g^{-1}\A^{-1}$ relating
$\omega_5(\ol{\cA})=\omega_5(\cA^{\A g^{-1}\A^{-1}})$
and $\omega_5(\cA)$, eq.\ \eqref{eq:rederivI} is rewritten into
\begin{align}
&\int_{\Msix}\tr\left(\cF^3 -(\cF^\cl)^3\right)
=\int_{\p\Msixzero=\Mfivezero+\Dtwo\times B}\left[
\left(\omega_5(\cA)-\omega_5(\ol{\cA})\right)
-\left(\omega_5(\cA^\cl)-\omega_5(\ol{\cA}^\cl)\right)\right]
\nn\\
&=-\int_{\p\Msixzero=\Mfivezero+\Dtwo\times B}\left\{
\frac{1}{10}\tr\!\left[L^5-\left(-igdg^{-1}\right)^5\right]
+d\!\left[\alpha_4(L,\cA)-\alpha_4(-igdg^{-1},\cA^\cl)\right]\right\}
\ ,
\label{eq:int_DtrF^3_NL}
\end{align}
where $L$ is given by
\begin{align}
L&=-i\A g\A^{-1}d\left(\A g^{-1}\A^{-1}\right)
\nn\\
&=-i\A\left[g(\A^{-1}d\A)g^{-1}-\A^{-1}d\A+ gdg^{-1}\right]\A^{-1}\ .
\label{eq:L}
\end{align}
Precisely speaking, $g=g(x;X^M)$ explicitly written in
\eqref{eq:int_DtrF^3_NL} and that appearing in  $L$ \eqref{eq:L} are
different ones: the former is from the classical solution and has a
constant and arbitrary instanton position $X^M$, while the latter has
$(t,s)$-dependent position $X^M(t,s)$.
However, we do not need to distinguish the two since the instanton
position can be absorbed by the shift of $x^M$ (note that the origin
$\xi=0$, the infinity $\xi=\infty$ and the boundary $B$ are defined in
terms of the relative coordinate $(x-X)^M$.)

First, let us confirm that we can safely discard the exact term
$d\bigl[\alpha_4(L,\cA)-\alpha_4(-igdg^{-1}\!\!,\cA^\cl)\bigr]$
in \eqref{eq:int_DtrF^3_NL}. A possible dangerous term at the origin
$\xi=0$ on $\Mfivezero$ is $\tr L^3\cA$ with
$L\Rightarrow\A(-ig\p_M g^{-1}dx^M)\A^{-1}\sim 1/\xi$ and
$\cA\Rightarrow\cA_0\,dt\sim \xi^0$. Taking this into account
and putting the boundary $\p\Mfourzero$ of infinitesimal radius
$\xi=\epsilon$, we have
\begin{align}
&\int_{\Mfivezero+\Dtwo\times B}
d\!\left[\alpha_4(L,\cA)-\alpha_4(-igdg^{-1},\cA^\cl)\right]
\nn\\
&=\int \! dt\!\int_{\p\Mfourzero}
\tr\left\{\left(\A(-ig\p_M g^{-1}dx^M)\A^{-1}\right)^3\cA_0
-(-ig\p_M g^{-1}dx^M)^3\cA^\cl_0\right\}
\nn\\
&=\int\! dt\!\int\!d\Omega_3\tr\left[
t_8\left(\Phi +i\A^{-1}\dot{\A}\right)\right]=0 \ ,
\label{eq:intd(alpha-alpha)}
\end{align}
where we have used $\A^{-1}\cA_0\A=\cA^\cl_0 +\Phi +i\A^{-1}\dot{\A}$
obtained from \eqref{eq:cptU(3)}, and
$(-ig\p_M g^{-1}dx^M)^3\linebreak \sim (1/\xi)^3\cP_2\,\xi^3d\Omega_3$
with $\cP_2$ given by \eqref{eq:cP_2}.
The last equality leading to zero is due to \eqref{sol:Phi} and that
$g(x)$ commutes with $t_8$.

Thus, we are left with the first term of
\eqref{eq:int_DtrF^3_NL}. The integrand can in fact be rewritten into
an exact form:
\begin{equation}
-\frac{1}{10}\tr\!\left[L^5-\left(-igdg^{-1}\right)^5\right]
=d\tr\left(\cO_A +\cO_B\right) \ ,
\end{equation}
with $\cO_A$ and $\cO_B$ given respectively by
\begin{align}
\cO_A&=-\frac{i}{2}
(-i\A^{-1}d\A)\left[(-igdg^{-1})^3-(-ig^{-1}dg)^3\right]\ ,
\\
\cO_B&=-\frac{i}{2}(-i\A^{-1}d\A)\Bigl[
g(-i\A^{-1}d\A)g^{-1}(-igdg^{-1})^2
-g^{-1}(-i\A^{-1}d\A)g(-ig^{-1}dg)^2
\nn\\
&\qquad\quad
-\frac12(-igdg^{-1})(-i\A^{-1}d\A)(-igdg^{-1})
+\frac12(-ig^{-1}dg)(-i\A^{-1}d\A)(-ig^{-1}dg)\Bigr] \ .
\end{align}
The $\cO_B$ term containing only two $-igdg^{-1}$ can be safely
dropped. However, the $\cO_A$ term with $(-igdg^{-1})^3\sim 1/\xi^3$
near the origin $\xi=0$ needs careful treatment like
\eqref{eq:intd(alpha-alpha)}.
Another way to evaluate
$\int_{\Mfivezero+\Dtwo\times B} d\cO_A$ is to note that
$d\cO_A=0$ holds on $\Mfivezero$ since we have
\begin{equation}
\bigl(-ig\p_M g^{-1}dx^M\bigr)^4
=\bigl(-ig^{-1}\p_M g dx^M\bigr)^4=0\quad
\mbox{on $\Mfour$} \ ,
\end{equation}
for a spherically symmetric $g(x)$ of \eqref{eq:g} (c.f.,
\eqref{eq:(R^cldx)^4=0}), and $(-i\A^{-1}d\A)^2=0$ on $\Mfive$
which is a surface with $s=0$.
Using this fact, we obtain
\begin{align}
\int_{\Mfivezero+\Dtwo\times B}d\tr\cO_A
&=\int_{\{s=0\}\times B}\tr\cO_A
=\tr\left\{\int\! dt\bigl(-i\A(t)^{-1}\dot{\A}(t)\bigr)
\,(-i)\!\int_{B=S^3}(-igdg^{-1})^3\right\}
\nn\\
&=\frac{24\pi^2}{\sqrt{3}}\int\! dt\,\tr\left[t_8
\bigl(-i\A(t)^{-1}\dot{\A}(t)\bigr)\right] \ ,
\end{align}
where we have used \eqref{eq:int(-igdg^-1)^3=} and \eqref{eq:cP_2}.
This implies our previous result \eqref{eq:evalnewSCS}.

\section{WZW term from $\bm{\newSCS}$}
\label{app:WZWfromnewSCS}

In this appendix, we see how the WZW term is correctly reproduced
from our new CS term \eqref{eq:newSCS} in the low energy limit.
We start with a configuration $\cA(t,x,s)$ in $\Msix$ which vanish at
the infinity $\xi=\infty$, and therefore at $z=\pm\infty$
(this configuration is not necessarily a baryon configuration).
For carrying out the expansion in terms of the modes in the $z$-space,
we move to the $\cA_z=0$ gauge via the gauge transformation by
\begin{equation}
V(t,\bm{x},z,s)=\Po\exp\left(
i\int_{-\infty}^z\! dz'\,\cA_z(t,\bm{x},z',s)\right) \ .
\end{equation}
The gauge field $\cA_\alpha$ ($\alpha=0,1,2,3,s$) in the
$\cA_z=0$ gauge satisfies the following
boundary condition (we use the same symbol $\cA$ for the gauge field
in the $\cA_z=0$),
\begin{equation}
\cA_\alpha(t,x,s)\to
\begin{cases}
-i\h(t,\bm{x},s)\p_\alpha\h(t,\bm{x},s)^{-1}
& (z\to +\infty)\\
0 & (z\to -\infty)
\end{cases}
\ ,
\end{equation}
with $\h(t,\bm{x},s)$ given by
\begin{equation}
\h(t,\bm{x},s)=\Po\exp\left(
i\int_{-\infty}^\infty\! dz\,\cA_z(t,\bm{x},z,s)\right) \ .
\label{eq:defh}
\end{equation}
Therefore, we can mode expand $\cA_\alpha$ as
\begin{equation}
\cA_\alpha(t,x,s)=-i\h(t,\bm{x},s)\p_\alpha\h(t,\bm{x},s)^{-1}
\times\psi_+(z) +\mbox{(massive modes)},
\label{eq:A_alpha=-ihph^-1}
\end{equation}
where $\psi_+(z)$ is the zero-mode given in \cite{SaSu1}:
\begin{equation}
\psi_+(z)=\frac12+\frac{1}{\pi}\arctan z
\to
\begin{cases}
1 & (z\to +\infty)\\
0 & (z\to -\infty)
\end{cases}
\ .
\end{equation}

Then, let us calculate our CS term \eqref{eq:newSCS} for the gauge
field \eqref{eq:A_alpha=-ihph^-1} by discarding the contribution from
the massive modes. First, the field strengths are given by
\begin{align}
\cF_{\alpha\beta}&=\p_\alpha\cA_\beta-\p_\beta\cA_\alpha
+i\left[\cA_\alpha,\cA_\beta\right]
\nn\\
&=i\left[-i\h\p_\alpha\h^{-1},-i\h\p_\beta\h^{-1}\right]
\psi_+(\psi_+ -1)
+\mbox{(massive modes)},
\nn\\
\cF_{z\alpha}&=\p_z\cA_\alpha
=-i\h\p_\alpha\h^{-1}\times\drv{}{z}\psi_+(z)
+\mbox{(massive modes)},
\end{align}
and using this we obtain
\begin{align}
\tr\cF^3&=\frac{6}{2^3}\tr\bigl(
\cF_{\alpha\beta}\cF_{\gamma\delta}\cF_{z\kappa}\bigr)
\,\epsilon^{\alpha\beta\gamma\delta\kappa}d^6x
\nn\\
&=3 \tr(-i\h d\h^{-1})^5\times\left[\psi_+(z)(\psi_+(z)-1)\right]^2
\drv{\psi_+(z)}{z}\,dz+\mbox{(massive modes)} \ .
\label{eq:trcF^3_WZW}
\end{align}
The $z$-integration of \eqref{eq:trcF^3_WZW} is trivially carried out
and we finally get the desired result:
\begin{equation}
\newSCS=\frac{N_c}{240\pi^2}
\int\tr\bigl(-i\h(t,\bm{x},s) d\h(t,\bm{x},s)^{-1}\bigr)^5
+\mbox{(contribution from massive modes)} \ .
\label{eq:newWZW}
\end{equation}
Note that this WZW term is different from the WZW term of
\cite{SaSu1}, \eqref{eq:SWZW_SS} with \eqref{eq:L=-iUdU^-1} and
\eqref{eq:U=PexpintAz}, in the definition of the Skyrme field $U$ in
terms of $\cA_z$ and in that the extra fifth coordinate is $s$ in the
present WZW term, while it is $z$ in \eqref{eq:SWZW_SS}.
However, in the topologically trivial sector without baryons, these
two WZW terms are equivalent since they anyhow are determined by
the Skyrme field at the boundary, namely, by
$\Po\exp\left(i\int_{-\infty}^\infty\! dz\,\cA_z(t,\bm{x},z)\right)$.

Let us consider the Skyrme field \eqref{eq:defh} for our WZW
term in the baryon sector.
In the baryon sector, the gauge field $\ol{\cA}(t,x,s)$
\eqref{eq:olcA(t,x,s)} in the singular gauge satisfies the condition
$\ol{\cA}\to 0$ ($\xi\to\infty$).\setcounter{footnote}{0}\footnote{
On the other hand, $\cA(t,x,s)$ in the regular gauge does not vanish
at $\xi=\infty$ since $\Delta A_0(t,x)\nrightarrow 0$ as
$z\to -\infty$. See the footnote above eq.\ \eqref{eq:DeltaA_0to0}.
}
Adopting
\begin{equation}
\ol{\cA}_z(t,x,s)=\A(t,s)\ol{A}_z^\cl(x)\A(t,s)^{-1} \ ,
\end{equation}
as $\cA_z$ in \eqref{eq:defh} (we ignore other collective coordinates
than $\A$), we have
\begin{equation}
\h(t,\bm{x},s)=\A(t,s)\h^\cl(\bm{x})\A(t,s)^{-1} \ ,
\label{eq:h=Ah^clA^-1}
\end{equation}
with $\h^\cl$ defined by (c.f., \cite{AtMa})
\begin{equation}
\h^\cl(\bm{x})=\Po\exp\left(
i\int_{-\infty}^\infty\! dz\,\ol{A}^\cl_z(\bm{x},z)\right) \ .
\end{equation}
Plugging \eqref{eq:h=Ah^clA^-1} with $s$-dependent $\h^\cl$
into \eqref{eq:newWZW} also leads to the same result,
eq.\ \eqref{eq:evalnewSCS}, as of course it should.
Concrete expressions of the various quantities are
\begin{equation}
\ol{A}_z^\cl(x)=\left(
\frac{1}{\xi^2+\rho^2} -\frac{1}{\xi^2}\right)
(\bm{x}\cdot\bm{\tau}) \ ,
\end{equation}
and
\begin{equation}
\h^\cl(\bm{x})
=\exp\Bigl(-i\ol{H}(r)\,\wh{\bm{x}}\cdot\bm{\tau}\Bigr) \ ,
\end{equation}
with
\begin{equation}
\ol{H}(r)=\pi\left(1-\frac{r}{\sqrt{r^2+\rho^2}}\right) \ .
\end{equation}
Note that $\ol{H}(r)$ has the same behavior as that of the
corresponding function of the Hedgehog solution in the Skyrme model
\cite{AtMa}; $\ol{H}(r=0)=\pi$ and $\ol{H}(r\to\infty)=\cO(1/r^2)$.
In any case, what is important for reproducing \eqref{eq:evalnewSCS}
is that the collective coordinate of the $SU(3)$ rotation, $\A$,
depends on the extra coordinate of the WZW term as well as on $t$.
This is not satisfied in \eqref{eq:U=AUclA^-1} where the extra
coordinate is $z$.


\end{document}